\documentclass[prd,aps,twocolumn,showkeys,nofootinbib, amsmath,amssymb]{revtex4-2}

\usepackage{amsmath}
\usepackage{amssymb}
\usepackage{array}
\usepackage{makecell}
\usepackage{xcolor}
\usepackage{acronym}
\usepackage{graphicx}
\usepackage{dcolumn}
\usepackage{bm}
\usepackage{soul}
\soulregister\cite7
\soulregister\ref7
\soulregister\ac7
\usepackage{hyperref}
\begin{document}

\title{Persistence of post-Newtonian amplitude structure in binary black hole mergers}

\author{Viviana A. Cáceres-Barbosa}
\affiliation{
 Institute for Gravitation and the Cosmos, Department of Physics, The Pennsylvania State University, University Park, Pennsylvania 16802, USA
}
\email{vac5288@psu.edu}

\date{\today}

\begin{abstract}
We analyze the spherical harmonic mode amplitudes of quasicircular, nonprecessing binary black hole mergers using 275 numerical relativity simulations from the SXS, RIT, and MAYA catalogs. We construct fits using the leading-order post-Newtonian (PN) dependence on intrinsic parameters, replacing the PN velocity with fit coefficients. We compare these to polynomial fits in symmetric mass ratio and spin. We analyze $(\ell, m)$ modes with $\ell \leq 4$ from late inspiral [$t = -500M$ relative to the $(2,2)$ peak] to postmerger ($t = 40M$). For nonspinning systems, the $(2,2)$, $(2,1)$, and $(3,3)$ modes retain the leading-order PN dependence on mass ratio throughout the merger. Higher-order modes deviate from the PN dependence only near and after the merger, where polynomial fits of degree $N \leq 3$ can capture the amplitude behavior up to $40M$. For aligned-spin systems at fixed mass ratio, the $(2,1)$ mode retains its PN spin dependence, while the $(3,2)$ and $(4,3)$ modes exhibit a quadratic spin dependence near merger. The PN-inspired fits lose accuracy with increasing mass ratio, particularly near merger. Results broadly agree across catalogs, though discrepancies appear in the $(3,1)$, $(4,2)$, and $(4,1)$ modes, likely from resolution differences. Our results clarify the extent to which PN structure persists in mode amplitudes. Although the fits cannot be fully interpreted within the PN formalism near merger, low-degree polynomial corrections to the PN amplitude Ans\"atze can capture strong-field behavior, enabling closed-form and efficient modeling of waveform amplitudes in this regime.
\end{abstract}

\maketitle

\acrodef{bbh}[BBH]{binary black hole}
\acrodef{gw}[GW]{gravitational-wave}
\acrodef{nr}[NR]{numerical relativity}
\acrodef{pn}[PN]{post-Newtonian}
\acrodef{sxs}[SXS]{Simulating Extreme Spacetimes}
\acrodef{rit}[RIT]{Rochester Institute of Technology}
\acrodef{bic}[BIC]{Bayesian Information Criterion}

\section{\label{sec:introduction}Introduction}
Recent advances in \ac{gw} detections \cite{soni2024ligo} have dramatically improved our ability to observe \ac{bbh} mergers, opening opportunities to study their dynamics, test general relativity, and describe the population of black hole binaries. The essential tool for these analyses is the ``waveform,'' which describes the time evolution of the strain that arrives at gravitational-wave detectors, such as The Laser Interferometer Gravitational-Wave Observatory \cite{aasi2015advanced}, Virgo \cite{acernese2014advanced}, and The Kamioka Gravitational Wave Detector \cite{akutsu2021overview}. 

The waveform of a \ac{bbh} system in a quasicircular orbit is characterized by eight intrinsic parameters: the two black hole masses and the six spin components. Efficiently extracting these parameters from an observed gravitational waveform and, conversely, constructing accurate waveforms for different intrinsic parameters helps maximize the scientific return of \ac{gw} observations.

The most precise method for constructing gravitational waveforms is by numerically solving Einstein's field equations \cite{lehner2014numerical}. However, \ac{nr} is computationally expensive and can only provide solutions for discrete points in the intrinsic parameter space. To overcome this limitation, approximation methods are necessary. 

During the early inspiral phase, when the two compact objects are widely separated and moving slowly, the \ac{pn} approximation provides an effective description of the waveform \cite{blanchet2024post}. However, as the binary approaches the merger, the \ac{pn} expansion breaks down due to the onset of strong-field effects and nonlinearities. In this regime, \ac{nr} is required to accurately model the dynamics and \ac{gw} emission. 

Spherical harmonics with spin weight $-2$ provide a basis for describing \ac{gw} radiation of inspiraling \ac{bbh}s at large distances from the source. Consequently, both \ac{pn} and \ac{nr} waveforms for binary signals are conventionally decomposed into modes using this basis \cite{blanchet2024post}. Searches and Bayesian inference of signal parameters can then selectively retain modes, balancing desired precision with computational cost.

In the ringdown phase, the \ac{gw} signal can be approximated by a perturbed Kerr solution as a superposition of quasinormal modes \cite{berti2009quasinormal}. The symmetries of the system during ringdown are best described by a decomposition into spheroidal harmonics, which correspond to the well-known quasinormal mode frequencies of a perturbed Kerr black hole. The angular indices specifying spheroidal harmonics are different from those specifying the spherical harmonics used for \ac{nr} waveforms, a phenomenon known as mode mixing \cite{kelly2013decoding}.  

In principle, the amplitudes and phases of spherical or spheroidal harmonic modes can be mapped back to the intrinsic parameters of the source. While the \ac{pn} expansion provides analytical expressions for this mapping in the inspiral regime, extracting similar relationships near the merger is more challenging. Nevertheless, \ac{bbh} detections dominated by these late-time regimes, like GW190521 \cite{abbott2020gw190521} and GW231123 \cite{LIGOScientific:2025rsn}, are already part of the observational record. Understanding and modeling the structure of mode amplitudes and phases in the strong-field regime across the vast \ac{bbh} parameter space will become increasingly important as more of these signals are detected.

Historically, the phase has received more attention in waveform modeling, primarily because matched filtering \cite{Owen:1998dk}, the main technique used for \ac{gw} detection and parameter estimation, is more sensitive to phase variations than amplitude. Consequently, phase evolution has been extensively studied in the literature. At the same time, several works have highlighted the importance of amplitude modeling, especially for tests of general relativity, and the inclusion of higher-order modes \cite{Tahura:2019dgr,Pan:2010hz,Pacilio:2024tdl}. The goal of this study is therefore to contribute to this line of work by focusing exclusively on mode amplitudes.

Previous studies \cite{borhanian2020comparison,kamaretsos2012black,kamaretsos2012hairloss,Islam:2024tcs,nobili2025ringdown,pacilio2024flexible, london2014modeling} have explored how mode amplitudes near merger and during the ringdown depend on the intrinsic parameters of the binary, both for spheroidal quasinormal modes in the ringdown and for spherical harmonic modes near merger. For instance, \cite{kamaretsos2012black} and \cite{kamaretsos2012hairloss} investigated the dependence of the quasinormal mode spectrum on the masses and spins of the progenitor system. Using \ac{nr} simulations from the \textsc{bam} \ac{nr} code \cite{brugmann2008calibration}, they obtained fitting functions that can describe the amplitudes of the dominant and loudest subdominant quasinormal modes in the ringdown phase. With a similar aim, the authors in \cite{nobili2025ringdown, pacilio2024flexible} used a Gaussian process regression to obtain mappings between intrinsic parameters and the ringdown quasinormal mode amplitudes for noneccentric precessing simulations from the \ac{sxs} catalog \cite{Scheel:2025jct, boyle2019sxs}. Eccentric simulations were considered in \cite{Carullo:2024smg}, where the author used simulations from the \ac{rit} catalog \cite{Healy:2017mvh, healy2020third} to obtain expressions for the ringdown mode amplitudes of nonspinning unequal-mass \ac{bbh}s.

On the other hand, Ref.~\cite{borhanian2020comparison} used simulations from the \ac{sxs} catalog to study the mode structure of spin-weighted spherical harmonics near the merger, using fits inspired by the leading-order dependence of the \ac{pn} mode amplitudes on binary parameters. Surprisingly, they found that \ac{pn}-inspired fits can describe mode amplitudes at fixed times near the merger, despite the breakdown of the \ac{pn} approximation in this region. The relationship between \ac{pn} and \ac{nr} waveforms was also studied in Ref.~\cite{Islam:2024tcs} for eccentric \ac{bbh} mergers using \ac{sxs}, \ac{rit}, and MAYA \cite{jani2016georgia,ferguson2023second} simulations. The author similarly found that \ac{pn} approximations remain valuable for constructing waveform models even close to the merger. However, discrepancies between the catalogs emerged as the mass ratio or eccentricity increased. 

Inspired by previous studies suggesting that specific modes retain the leading-order \ac{pn} dependence on intrinsic parameters close to the merger, we first investigate the stability of leading-order \ac{pn} fits for spherical harmonic mode amplitudes from the late inspiral through the postmerger regime. We connect the early-time behavior of the recovered fit coefficients to the \ac{pn} velocity to aid in the interpretation of the recovered values.

Second, we construct polynomial fits in mass ratio and spins to assess the role of beyond-leading-order dependencies in improving amplitude fits, particularly for modes where leading-order fits break down near merger. We aim to identify the simplest model required to capture the dominant physical dependencies for each mode.

Finally, for the first time, we use these methods to carry out a detailed comparison of \ac{nr} simulations from the \ac{sxs}, \ac{rit}, and MAYA \ac{nr} catalogs. Cross-catalog comparisons provide a critical consistency check and help flag simulations where systematic differences may emerge.

The rest of this paper is organized as follows. Section~\ref{sec:methods_spherical_harmonics} provides an overview of the spherical harmonic decomposition and introduces relevant quantities and notation. Section~\ref{sec:methods_NR_catalogs} briefly describes the different \ac{nr} catalogs used in this study and outlines the simulation selection criteria. Sections~\ref{sec:methods_PN_Expansion} and \ref{sec:lo_methods} present the validity of the \ac{pn} expansion and the construction of \ac{pn}-inspired fits, along with the methods used to assess their performance and stability. Section~\ref{sec:ho_methods} introduces fits with beyond-leading-order parameter dependencies. Section~\ref{sec:results_lo} discusses the stability and evolution of the fit coefficients of the \ac{pn}-inspired fits, evaluates their performance, interprets their connection to the true \ac{pn} expansion at earlier times, and compares results across \ac{nr} catalogs. Section~\ref{sec:results_ho} presents the results of fits with higher-order dependencies and evaluates their improvements over the leading-order models. Finally, Sec.~\ref{sec:conclusions} summarizes our findings and outlines directions for future work.

We use a geometrized system of units in which Newton's gravitational constant $G$ and the speed of light $c$ are $c=G=1$. 

\section{\label{sec:methods}Methods}
\subsection{Decomposition of gravitational-wave strain into spherical harmonic modes}\label{sec:methods_spherical_harmonics}
The multipolar expansion of \ac{gw} radiation from \ac{bbh} mergers can be expressed in a basis of $-2$ spin-weighted spherical harmonics $Y_{\ell m}^{-2}(\theta, \phi)$ as follows:
\begin{gather}
    h(t;\theta, \phi) = \sum_{\ell=2}^{\infty}\sum_{m=-\ell}^{m=\ell} h_{\ell m}(t) \, Y_{\ell m}^{-2}(\theta, \phi), \label{eq:strain}
\end{gather}
where $h$ is the complex gravitational-wave strain, $h_{\ell m}$ are the wave modes, $(\theta, \phi)$ are the angular coordinates specifying the direction of wave propagation, and $t$ is the retarded time  \cite{thorne1980multipole}. 

Each complex $(\ell, m)$ mode can be written in polar form as
\begin{gather}
    \displaystyle h_{\ell m}(t) = A_{\ell m}(t)e^{i\phi_{\ell m}(t)},
\end{gather}
where $A_{\ell m}(t)$ and $\phi_{\ell m}(t)$ are its amplitude and phase, respectively. The $(2, 2)$ mode dominates the gravitational radiation emitted by a binary system \cite{kidder2008using, mishra2016ready}. In this study, we analyze the amplitudes of different modes relative to the $(2, 2)$ mode,
\begin{gather*}
    \hat{A}_{\ell m}(t) = \begin{cases}
        A_{\ell m}(t) & \text{if }(\ell, m) = (2, 2)\\
        \displaystyle\frac{A_{\ell m} (t)}{A_{22}(t) }& \text{otherwise}.
    \end{cases}
\end{gather*}

\begin{figure}
    \includegraphics[width=\linewidth]{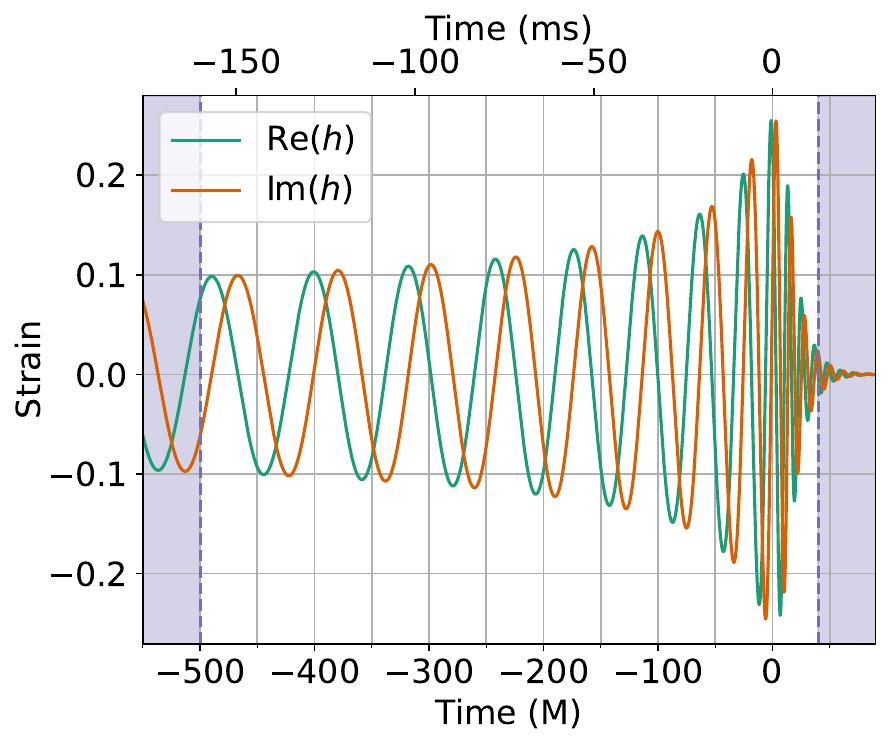}
    \caption{\label{fig:timescales}Real and imaginary parts of the 
    \ac{gw} strain $h$, evaluated in the direction $(\theta, \phi) = (0, 0)$ for the simulation SXS:BBH:0305. The total mass is set to be $M=65M_\odot$. The lower horizontal axis shows time relative to the peak of the $(2, 2)$ mode in units of $M$, while the upper horizontal axis converts to milliseconds. The purple shaded region marks times outside the interval $t \in (-500M, 40M)$, which are excluded from this work.}
\end{figure}

The $\hat{A}_{\ell m}(t)$ are extracted over the interval $t \in (-500M, 40M)$, where $t=0M$ corresponds to the peak amplitude of the $(2, 2)$ mode.\footnote{\normalfont The total mass $M$ of the binary sets the characteristic timescale of the system. In natural units, where $c=1$ and $G=1$, time can be expressed in terms of $M$, with $ 1M_\odot \approx 5 \times 10^{-6}$ s.} Figure~\ref{fig:timescales} illustrates this interval on the real and imaginary parts of the full strain $h$ [Eq.~(\ref{eq:strain})] for the simulation SXS:BBH:0305, which appears in Fig.~1 of the GW150914 discovery paper \cite{abbott2016observation}. We highlight the situation of this interval near and around the merger. 

The primary dynamical parameter for a nonspinning system is its mass ratio $q = \frac{m_1}{m_2}$.  Using the convention $ m_1 \geq m_2$, such that $q \geq 1$, we can also define the symmetric mass ratio $\eta$ and the mass asymmetry parameter $\delta$,
\begin{gather}
    \label{eq:mass_ratio_parameters}
    \eta \equiv \frac{m_1 m_2}{(m_1 + m_2)^2}, \\  
    \delta \equiv \frac{m_1 - m_2}{m_1 + m_2} = \sqrt{1 - 4\eta} \, .
\end{gather}
In aligned-spin binaries, where the spin vectors are parallel to the orbital angular momentum, the waveform is determined by $q$ and the spin magnitudes of the two bodies, given in dimensionless form as $\chi_i = \frac{S_i}{m_i^2}$, with $i=1,2$ labeling the object in the binary. It is also useful to define the symmetric and antisymmetric spin combinations,
\begin{gather}
    \label{eq:sym_antisym_spins}
    \chi_s \equiv \frac{1}{2}\left( \chi_1 + \chi_2 \right), \\
    \chi_a \equiv \frac{1}{2}\left( \chi_1 - \chi_2 \right).
\end{gather}

\subsection{Numerical relativity simulations}\label{sec:methods_NR_catalogs}
Several \ac{nr} groups have performed simulations of \ac{bbh} mergers using different formalisms to set initial conditions and evolve the spacetime metric \cite{Scheel:2025jct, boyle2019sxs, Healy:2017mvh, healy2020third,jani2016georgia,ferguson2023second, husa2008reducing, brugmann2008calibration}. In principle, simulations with identical initial conditions should produce the same gravitational waveform. However, differences in numerical precision, gauge choices, and waveform extrapolation techniques, among other factors, can introduce discrepancies between catalogs \cite{lovelace2016modeling}. 

To compare results across catalogs, we use simulations from the \ac{sxs} \cite{Scheel:2025jct, boyle2019sxs}, \ac{rit} \cite{Healy:2017mvh, healy2020third}, and MAYA \cite{jani2016georgia,ferguson2023second} catalogs. In this section, we briefly review the methods employed by each catalog and outline our simulation selection criteria.

\subsubsection{SXS catalog}
The \ac{sxs} catalog utilizes the spectral Einstein code \cite{blackholesSpECSpectral} to perform simulations. The initial data are generated by solving the extended conformal thin-sandwich equations, which are discretized on a grid and solved using a spectral elliptic solver. Iterative tuning is applied to achieve the desired \ac{bbh} properties, and the initial data are evolved for several orbits to reduce eccentricity and produce quasicircular orbits \cite{boyle2019sxs}.

The evolution of the initial data is carried out using a first-order version of the generalized harmonic formulation \cite{friedrich1985hyperbolicity, garfinkle2002harmonic, pretorius2005numerical} of Einstein's equations, with constraint damping. A multidomain spectral method is applied, where the computational domain is partitioned into subdomains at various stages. A spectral adaptive mesh refinement algorithm determines when grid points should be added or subtracted in a given subdomain, while a control system and the associated maps shift the boundaries of the various subdomains with respect to each other in order to accommodate the requirement that the excision boundary must stay inside the apparent horizons at all times \cite{Hemberger:2012jz, Szilagyi:2014fna}. These techniques rely on a set of specified error tolerances that are determined by the resolution of a simulation \cite{Scheel:2025jct}.

In this work, we use the highest available resolution level for each \ac{sxs} simulation. The waveforms were read using the \textsc{sxs python} package \cite{SXSPackage_v2025.0.15, SXSCatalogData_3.0.0}.  

\subsubsection{RIT catalog}
The simulations in the \ac{rit} catalog are generated using the \textsc{LazEv} code \cite{zlochower2005accurate}. The initial data for \ac{rit} simulations are derived from the Bowen-York solution and computed using a generalized version of the \textsc{TwoPunctures} code \cite{ansorg2004single}, which solves a coupled system of the Hamiltonian and momentum constraints. \ac{rit} simulations use \ac{pn} approximations \cite{Healy:2017zqj} to produce quasicircular orbits.

The \textsc{LazEv} code implements the Baumgarte-Shapiro-Shibata-Nakamura-Oohara-Kojima \cite{baumgarte1998numerical,shibata1995evolution, nakamura1987general} formulation of Einstein’s equations. The code utilizes the \textsc{cactus} \cite{Goodale2002a} infrastructure from the \textsc{EinsteinToolkit} \cite{loffler2012einstein} and uses the \textsc{carpet} \cite{schnetter2004evolutions} module for mesh refinement. Each simulation is labeled nXYY, where the grid spacing in the wave zone is specified as $M/X.YY$.  For example, n120 corresponds to a grid spacing of $M/1.2$ in the wave zone.

Only simulations with a resolution label of n120 or higher are included from the \ac{rit} catalog. The simulations were read directly from publicly available strain data files \cite{CCRGNumericalSimulations}.

\subsubsection{MAYA catalog}
The simulations in the MAYA catalog use the \textsc{maya} code \cite{ferguson2023second}. Similar to \textsc{LazEv}, the \textsc{maya} code uses the \textsc{cactus}/\textsc{carpet}/\textsc{EinsteinToolkit} infrastructure for its simulations. It constructs initial data based on the Bowen-York extrinsic curvature, conformally flat spatial metric, and the \textsc{TwoPunctures} solver. The \ac{pn} equations of motion are also used to define the initial spins and orbital momentum. The \textsc{maya} code uses the Baumgarte-Shapiro-Shibata-Nakamura  \cite{baumgarte1998numerical,shibata1995evolution} formulation to evolve the initial data, employing mesh refinement for higher-resolution simulations.

The publicly available MAYA data files \cite{UTAustinWaveforms} were read using the \textsc{mayawaves} \cite{Ferguson_mayawaves_2023} \textsc{python} package.

\subsubsection{Selection criteria}
The linear shift in the time origin of the coordinates, which arises from the residual motion of the center of mass of the binary system \cite{woodford2019compact}, was corrected for all simulations used. Additionally, all simulations are selected to have initial orbital eccentricities of less than 0.002, as reported by their respective catalogs.\footnote{Eccentricity is not consistently defined across catalogs. While there are methods \cite{Shaikh:2025tae} for consistently calculating eccentricity using waveform data, reevaluating the eccentricity for all simulations considered here is beyond the scope of this work.}

For nonspinning binary systems, simulations with any mass ratio are included, provided that the initial magnitude of each of the six dimensionless spin components is less than $10^{-4}$. Table \ref{tab:nonspinning_all} in Appendix~\ref{app:simulations_info} lists the nonspinning simulations used in this work.

For aligned-spin binary systems, simulations with fixed mass ratios $q \in \{1.0, 1.5, 2.0\}$ are used. The initial perpendicular components of the dimensionless spins are required to be less than $10^{-4}$. Additionally, $\chi_i \geq 0.1$ for at least one of the black holes to enhance the impact of spins on the waveforms. Table \ref{tab:alignedspin_all} lists the aligned-spin simulations used from the \ac{sxs}, \ac{rit}, and MAYA catalogs.

We refer to the selected simulations as ``datasets'' for each \ac{nr} catalog. After extracting the mode amplitudes $\hat{A}_{\ell m}$ from each dataset, their dependence on the binary parameters was modeled using fits motivated by the leading-order \ac{pn} dependence.

\subsection{Post-Newtonian expansion}\label{sec:methods_PN_Expansion}
The \ac{pn} approximation describes the motion of inspiraling compact binaries and their emission of \ac{gw}s \cite{blanchet2024post}. It is an expansion in the characteristic internal velocity $v$ of the system, making the approximation valid for sources that are slowly moving and weakly stressed. Several works \cite{blanchet2024post, kidder2008using, arun2009higher, henry2022spin} have computed the complex $h_{\ell m}$ modes at various \ac{pn} expansion orders. The expressions for the relative mode amplitudes $\hat{A}_{\ell m}$ referenced throughout this work are shown in Appendix~\ref{app:pn_expressions} and can be obtained from Refs.~\cite{blanchet2024post, kidder2008using, henry2022spin, pan2014inspiral, arun2009higher}.

For nonspinning binaries, we focus on modes with $\ell \leq 4$ and $0 < m \leq \ell$.\footnote{Nonprecessing systems exhibit mirror symmetry, meaning that the $-m$ modes are fully determined by the positive-$m$ modes through the relation $h_{\ell,-m} = (-1)^{\ell} h_{\ell m}^*$. For this reason, we focus only on modes with $m > 0$.} Among these, the $(2,2)$, $(2,1)$, $(3,3)$, and $(4,4)$ modes are generally the strongest \cite{mills2021measuring}.

For aligned-spin systems, we consider the modes $\{ (\ell, m) \} = \{(2, 2), (2, 1), (3, 2), (4, 3), (4, 1)\}$. We include the dominant $(2, 2)$ mode due to its central role in waveform modeling, while the remaining modes are analyzed because they correspond to current multipoles. Modes with $\ell + m = \mathrm{odd}$ are entirely given by the current multipole moment, which is directly related to the system's total angular momentum \cite{blanchet2024post}. Hence, while spin effects are not present at leading order for any mode, these current multipole modes contain spin-dependent terms at next-to-leading order, making them more suitable for probing spin effects.

\subsection{Investigating the leading-order post-Newtonian dependence} \label{sec:lo_methods}
We examine how the \ac{gw} mode amplitudes vary with the mass ratio and spins of the system over a range of times using fit Ans\"atze that are motivated by \ac{pn} theory. Near the merger, the \ac{pn} approximation breaks down. However, Ref. \cite{borhanian2020comparison} suggested that the leading-order dependence of the mode amplitudes on the intrinsic parameters persists through the merger. The first objective of this study is to derive and reinterpret these results by directly fitting for $v$ in the \ac{pn} expressions at each time step. 

Following Ref.~\cite{borhanian2020comparison}, a fit is constructed for each mode based on the leading-order functional form of the \ac{pn} expressions (see Appendix \ref{app:pn_expressions}). Since this previous work found good agreement between their fits and the \ac{nr} data at $-100M$, we expect $-500M$ to be a conservative choice for start time. Unlike Ref.~\cite{borhanian2020comparison}, we look for correlations across modes by directly replacing powers of $v$ with independent fit coefficients instead of fitting for an overall factor. The fit coefficient that replaces the $v^n$ term in $\hat{A}_{\ell m}$ is denoted $a_{\ell m}^{(n)}$. We also define $a_{\ell m} \equiv \sqrt[n]{a_{\ell m}^{(n)}}$.

\subsubsection{Leading-order fits for nonspinning simulations}
For nonspinning simulations, the resulting leading-order fitting functions are
\begin{subequations} \label{eq:nonspinning_lo_fits}
\begin{align}
    \hat{A}_{22}^{(\mathrm{LO})} &= 8 \sqrt{\frac{\pi}{5}} \, \eta \, a_{22}^{(2)}, \label{eq:nonspinning_lo_fits_22}\\
    \hat{A}_{21}^{(\mathrm{LO})} &= \frac{1}{3} \, \delta \, a_{21}^{(1)}, \label{eq:21_nonspinning_fit}\\
    \hat{A}_{33}^{(\mathrm{LO})} &= \frac{3}{4} \sqrt{\frac{15}{14}} \, \delta \, a_{33}^{(1)}, \\
    \hat{A}_{32}^{(\mathrm{LO})} &= \frac{1}{3}\sqrt{\frac{5}{7}}\,(1 - 3\eta)\,a_{32}^{(2)}, \\
    \hat{A}_{31}^{(\mathrm{LO})} &= \frac{\delta}{12\sqrt{14}}a_{31}^{(1)},\\
    \hat{A}_{44}^{(\mathrm{LO})} &= -\frac{8}{9} \sqrt{\frac{5}{7}} (3 \eta - 1) a_{44}^{(2)}, \\
    \hat{A}_{43}^{(\mathrm{LO})} &= \frac{9\,\delta\, (1 - 2\eta)}{4\sqrt{70}}a_{43}^{(3)}, \\
    \hat{A}_{42}^{(\mathrm{LO})} &= \frac{\sqrt{5}}{63}\,(1 - 3\eta)\,a_{42}^{(2)}, \\
    \hat{A}_{41}^{(\mathrm{LO})} &= \frac{\delta\,(1 - 2\eta)}{84\sqrt{10}}a_{41}^{(3)}.
\end{align}
\end{subequations}

To perform the fits, we extract $\hat{A}_{\ell m}$ from the \ac{nr} simulation at each time step and apply the fitting functions presented above. This procedure yields a set of time-dependent fit coefficients $a_{\ell m}^{(n)}(t)$, which enable us to track their evolution and physical significance.

For nonspinning simulations, we apply the fits to the individual catalogs and to the joint dataset to assess their consistency.

One might expect the $a_{\ell m}(t)$ to agree across modes at early times, when the \ac{pn} expansion is more valid, because they replace $v$ in the \ac{pn} expressions, which is common to all modes. However, unless the amplitudes are taken early enough, such that the \ac{pn} expansion can be truncated at leading order for all modes to good approximation, the fit coefficients need not agree between modes. This is because, on one hand, the velocity itself depends on the mass ratio and spins, and on the other hand, \ac{pn} corrections beyond leading order can still be relevant at early times. As an illustration, if we write out dependencies on $t$ and $\eta$ explicitly, the $(2,2)$ mode is well approximated in the \ac{pn} regime by Eq.~\eqref{eq:nonspinning_lo_fits_22} if
\begin{align}
    8 \sqrt{\frac{\pi }{5}} \eta  \left[v(t; \eta)\right]^2 + \dots &\approx 8 \sqrt{\frac{\pi}{5}} \, \eta \, a_{22}^{(2)}(t).
\end{align}

Therefore, $a_{22}^{(2)}(t)$ absorbs components of $\left[v(t; \eta)\right]^2$ that do not depend on $\eta$ as well as components of higher-order corrections to the amplitude that carry a linear $\eta$ dependence, like the leading-order term. We explore the relationship between $v$ and the $a_{\ell m}$ further by applying the fits at $t=-8000M$ for the subset of nonspinning \ac{sxs} simulations that begin earlier than this time before merger, well into the early inspiral regime.

\subsubsection{Leading-order fits for aligned-spin simulations}
For aligned-spin systems, the resulting fits are
\begin{subequations} \label{eq:alignedspin_lo_fits}
\begin{align}
    \hat{A}_{22}^{(\mathrm{LO, S})} &= \left| \hat{A}_{22}^{\mathrm{ns}}+ \frac{32}{3} \sqrt{\frac{\pi }{5}} \eta \left[(\eta -1) \chi_s-\delta  \chi _a\right] a_{22}^{(5)}\right|, \\
    \hat{A}_{21}^{(\mathrm{LO, S})} &= \left| \hat{A}_{21}^{\mathrm{ns}}-\frac{1}{2} \left(\chi _a+\delta  \chi _s\right)a_{21}^{(2)} \right|, \\
    \hat{A}_{32}^{(\mathrm{LO, S})} &=  \left|\hat{A}_{32}^{\mathrm{ns}}+\frac{4}{3} \sqrt{\frac{5}{7}} \eta  \chi _s a_{32}^{(3)}\right|, \\
    \hat{A}_{43}^{(\mathrm{LO, S})} &= \left| \hat{A}_{43}^{\mathrm{ns}}+\frac{45}{8 \sqrt{70}} \eta  \left(\chi _a-\delta  \chi _s\right)  a_{43}^{(4)} \right|, \\
    \hat{A}_{41}^{(\mathrm{LO, S})} &=  \left| \hat{A}_{41}^{\mathrm{ns}} + \frac{5}{84 \sqrt{40}} \eta  \left(\delta  \chi _s-\chi _a\right) a_{41}^{(4)} \right|,
\end{align}
\end{subequations}
where $\hat{A}_{\ell m}^{\mathrm{ns}}$ is the $(\ell, m)$ mode amplitude of a nonspinning system with the corresponding mass ratio. The amplitudes are extracted at each time step from aligned-spin simulations with fixed mass ratios $q=\{1.0, 1.5, 2.0\}$. With the time and mass ratio fixed, the amplitudes are treated as functions of $(\chi_{1z}, \chi_{2z})$, and the fits above are applied. 

In Ref.~\cite{borhanian2020comparison}, the authors obtained results for aligned-spin systems separate from those for nonspinning systems. However, suppose the leading-order spin combination [e.g., $\chi_a + \delta \chi_s$ for the $(2, 1)$ mode] truly governs the mode's spin dependence through the merger. Whenever this effective spin combination vanishes, the amplitude should then closely match that of a nonspinning system with the same mass ratio. Therefore, to fully isolate the spin dependence, we set $\hat{A}_{\ell m}^{\mathrm{ns}}$ in the above fits to the \ac{nr} amplitudes from SXS:BBH:4434 when $q = 1.0$, SXS:BBH:3984 when $q = 1.5$, and SXS:BBH:2497 when $q = 2.0$. Note that odd-$m$ modes are not excited for equal-mass nonspinning systems, and there is liberty to pick the overall sign of the fit coefficients for systems with $q=1.0$.

\subsection{Introducing higher-order dependencies}\label{sec:ho_methods}
In previous work, the amplitudes of several modes, most notably the $(3,2)$ mode, but also the $(3,1)$, $(4,3)$, $(4,2)$, and $(4,1)$ modes, exhibited deviations from the leading-order \ac{pn} fits at late times \cite{borhanian2020comparison}. The second objective of this study is to develop improved models for the relative mode amplitudes by incorporating higher-order dependencies on the intrinsic parameters.

\subsubsection{Higher-order fits for nonspinning simulations}
We construct fitting functions for the relative mode amplitudes of nonspinning systems that incorporate higher-order dependencies on the symmetric mass ratio $\eta$. Our primary focus in this section is on the quality of the fit, rather than the physical interpretation of the coefficients. Therefore, we do not explicitly replace powers of $v$ with a fit coefficient. Instead, we allow the coefficients to absorb all overall factors and velocity dependencies, and we determine which modes and times are better described with these higher-order fits.

By inspection of the \ac{pn} expressions in Appendix~\ref{app:pn_expressions}, we observe that for modes with even $m$, higher-order \ac{pn} terms introduce higher powers of $\eta$. In modes with odd $m$, the amplitudes additionally acquire an overall factor of $\delta$. Thus, we define a family of higher-order fitting functions $\hat{A}_{\ell m}^{(N)}$ for nonspinning systems as follows:
\begin{align}\label{eq:nonspinning_ho_fits}
    \hat{A}_{\ell m}^{(N)} &=
    \begin{cases}
        \displaystyle\sum_{i=0}^N c_i \eta^i, & \text{for even } m \\
        \delta\,\displaystyle\sum_{i=0}^N c_i \eta^i, & \text{for odd } m.
    \end{cases}
\end{align}
These fits allow us to determine the minimum degree of higher-order dependencies needed to model deviations from leading-order fits.

\subsubsection{Higher-order fits for aligned-spin simulations}
Similarly, for aligned-spin systems, we define higher-order fitting functions that incorporate linear and quadratic dependencies on $\chi_s$ and $\chi_a$,
\begin{subequations}\label{eq:alignedspin_ho_fits}
\begin{align}
    \hat{A}_{\ell m}^{(1,\text{S})} &= \hat{A}_{\ell m}^{\mathrm{ns}} + \sum_{i=a,s} c_i \chi_i, \label{eq:alignedspin_ho_linear}\\
    \hat{A}_{\ell m}^{(2,\text{S})} &= \hat{A}_{\ell m}^{\mathrm{ns}} + \sum_{i=a,s} c_i \chi_i + \sum_{i=a,s} \sum_{j=a,s}c_{ij} \chi_{i}\chi_{j}.\label{eq:alignedspin_ho_quadratic}
\end{align}
\end{subequations}
As done with the leading-order fits, we fix the mass ratio and use the nonspinning \ac{nr} reference amplitudes $\hat{A}_{\ell m}^{\mathrm{ns}}$.

\subsection{Fitting procedure}
The optimal fit coefficients are obtained using \textsc{SciPy}'s differential\_evolution algorithm. This global optimization method evolves a population of candidate solutions to minimize a given cost function \cite{2020SciPy-NMeth}. We set the cost function to the sum of squared residuals between the fit and the \ac{nr} data.

Although one might expect the fitted leading-order coefficients, which stand in for $v$, to lie between $0$ and $1$, these coefficients do not recover $v$ itself, as previously discussed. Since the leading-order coefficient absorbs higher-order corrections beyond the leading-order behavior, we allow it to vary over the broader range $[-5, 5]$. In practice, the coefficients typically take on values much smaller in magnitude than the imposed bounds, indicating that the results are not driven by this prior range.

Because we apply the fit at each time step independently, we take the smooth variation of fit coefficients, as well as consistency across catalogs, as evidence against overfitting, since these indicate that the coefficients capture systematic trends rather than noise.

For higher-order fits, we find that the coefficients may saturate the bounds $[-5, 5]$, and we therefore extend the allowed range to $[-10, 10]$. As before, the smooth variation of the recovered coefficients suggests that the fits are well behaved. However, allowing such a wide parameter range may increase the risk of overfitting or introduce multiple local minima in the optimization, particularly for higher-order models with a larger number of free parameters. To address these concerns, in Appendix \ref{app:bayesian} we outline and perform a full Bayesian regression analysis for the higher-order models near merger. We find that the results generally align with those of the differential evolution algorithm.

The quality of each fit is evaluated using the Pearson correlation coefficient,
\begin{gather}
    \label{eq:pearson_correlation}
    C_{\ell m} = \frac{(\hat{A}_{\ell m}^{(\mathrm{fit})} - \bar{A}_{\ell m}^{(\mathrm{fit})}) \cdot (\hat{A}_{\ell m}^{(\mathrm{NR})} - \bar{A}_{\ell m}^{(\mathrm{NR})})}{\sqrt{(\hat{A}_{\ell m}^{(\mathrm{fit})} - \bar{A}_{\ell m}^{(\mathrm{fit})})^2 (\hat{A}_{\ell m}^{(\mathrm{NR})} - \bar{A}_{\ell m}^{(\mathrm{NR})})^2}},
\end{gather}
where $\hat{A}_{\ell m}^{(\mathrm{fit})}$ represents the predicted amplitude values from the fitting functions, and $\hat{A}_{\ell m}^{(\mathrm{NR})}$ is the reference amplitude data obtained from \ac{nr} simulations. Overbars denote the mean values of the datasets.

\section{Results and Discussion}
\subsection{Relationship between fit coefficients and PN velocity}\label{sec:inspiral}
\begin{figure}
\includegraphics[width=\linewidth]{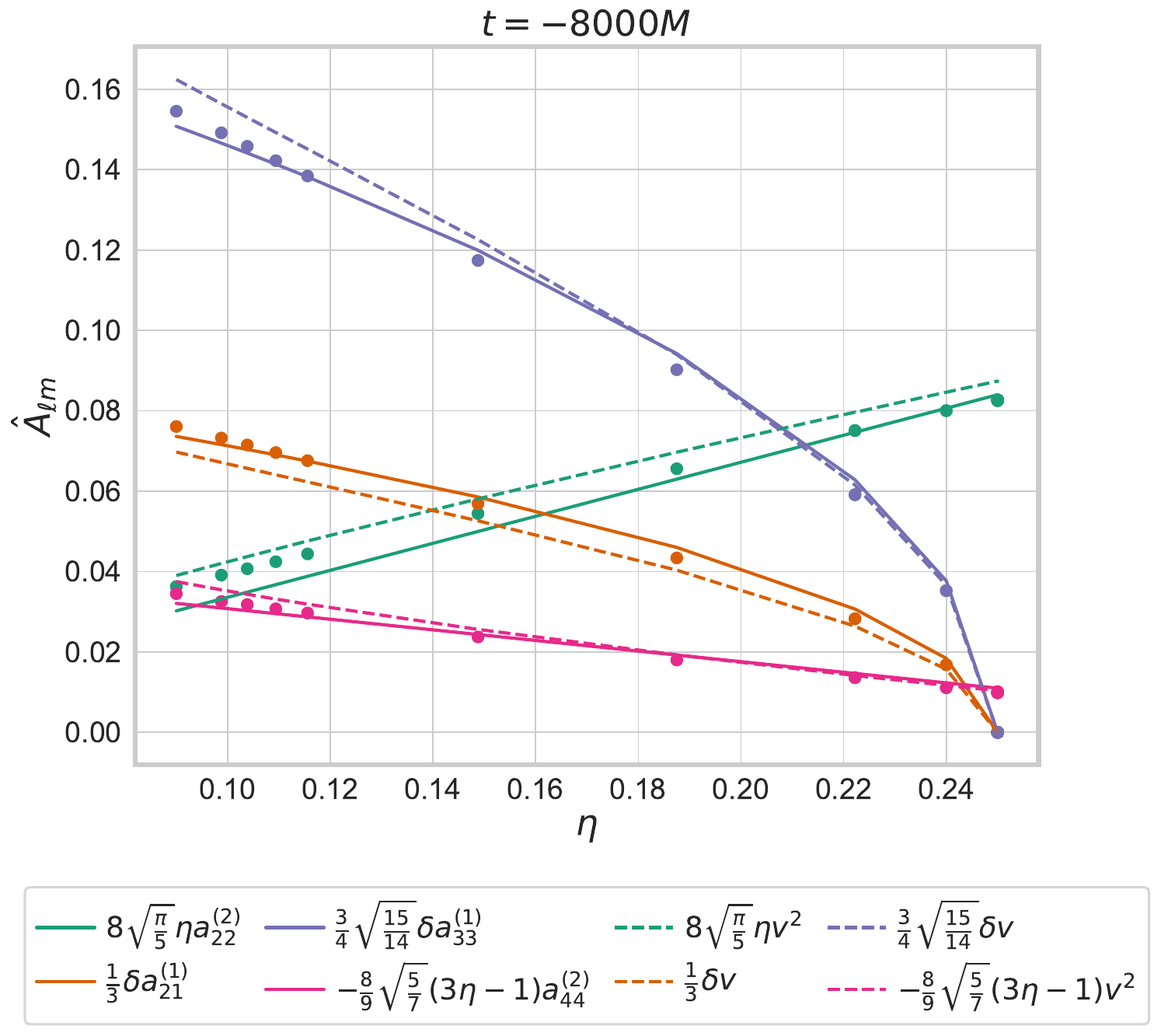}
    \caption{\label{fig:inspiral_amplitudes} The $(2, 2)$, $(2, 1)$, $(3, 3)$, and $(4, 4)$ mode amplitudes at $t=-8000M$ for the subset of nonspinning \ac{sxs} simulations that begin earlier than this time before merger. Solid curves show the best-fit curves obtained by fitting for the $a_{\ell m}$ in Eqs.~(\ref{eq:nonspinning_lo_fits}), while dotted curves show the leading-order \ac{pn} term obtained using $v$ from each simulation.}
\end{figure}

To illustrate the relationship between $v$ and the fit coefficients $a_{\ell m}$, Fig. \ref{fig:inspiral_amplitudes} shows the mode amplitudes at $t=-8000M$ for the strongest modes considered. We compare the best-fit curve obtained using Eqs.~(\ref{eq:nonspinning_lo_fits}) to the amplitudes obtained by truncating the \ac{pn} expressions at leading order. As can be seen in the figure, higher-order corrections are small but relevant at this early time, and the leading-order \ac{pn} term does not fully capture the behavior of the mode amplitudes. Furthermore, the best-fit curves are not equivalent to the leading-order \ac{pn} amplitudes. As previously mentioned, the fits absorb higher-order corrections, which may differ across modes. Therefore, the recovered fit coefficients, which are shown in Table \ref{tab:inspiral_fits}, are close but not equivalent. Note also that fitting for the $a_{\ell m}$ at a given time produces a single number independent of $\eta$, while $v$, in general, depends on $\eta$. Consequently, although the $a_{\ell m}$ stand in for $v$ in the \ac{pn} expressions and recover similar values at early times, they are not immediately comparable to a single reference value of $v$.

We note that the extrapolated \ac{nr} waveforms used in this work are not in the same Bondi-Metzner-Sachs (BMS) frame as \ac{pn} waveforms, and \ac{pn} parameters need not coincide with quasilocal mass and spin measurements at the apparent horizons that are quoted in \ac{nr} simulations. Recent work shows one can align the BMS frames when hybridizing \ac{nr} and \ac{pn} waveforms by optimizing over the \ac{pn} parameters to find the best fit to the \ac{nr} waveform \cite{Sun:2024kmv,Sun:2025una}. While this optimization over \ac{pn} parameters is beyond the scope of this study, we expect that systematic differences between \ac{pn} parameters and quasilocal \ac{nr} measurements are implicitly absorbed into the fit coefficients. Discrepancies in the fitted coefficients at early times may partially reflect this parameter mismatch.

\begin{table}[]
    \centering
    \begin{tabular}{c|c}
        Mode $(\ell, m)$ & $a_{\ell m}(t = -8000M)$ \\
        \hline
        $(2, 2)$ & $0.23017$ \\
        $(2, 1)$ & $0.27610$ \\
        $(3, 3)$ & $0.24277$ \\
        $(4, 4)$ & $0.24192$ \\
    \end{tabular}
    \caption{Optimal fit coefficients for the fits in Eqs.~(\ref{eq:nonspinning_lo_fits}) at $t=-8000M$. }
    \label{tab:inspiral_fits}
\end{table}

When applying the fits near merger, higher-order corrections may cause the optimal $a_{\ell m}^{(n)}$ to be negative. In the following sections, we quote $a_{\ell m}^{(n)}$ rather than $a_{\ell m}$.

\subsection{Stability and time evolution of leading-order PN-inspired fits}\label{sec:results_lo}
\subsubsection{Nonspinning simulations}
\begin{figure*}
\includegraphics[width=\linewidth]{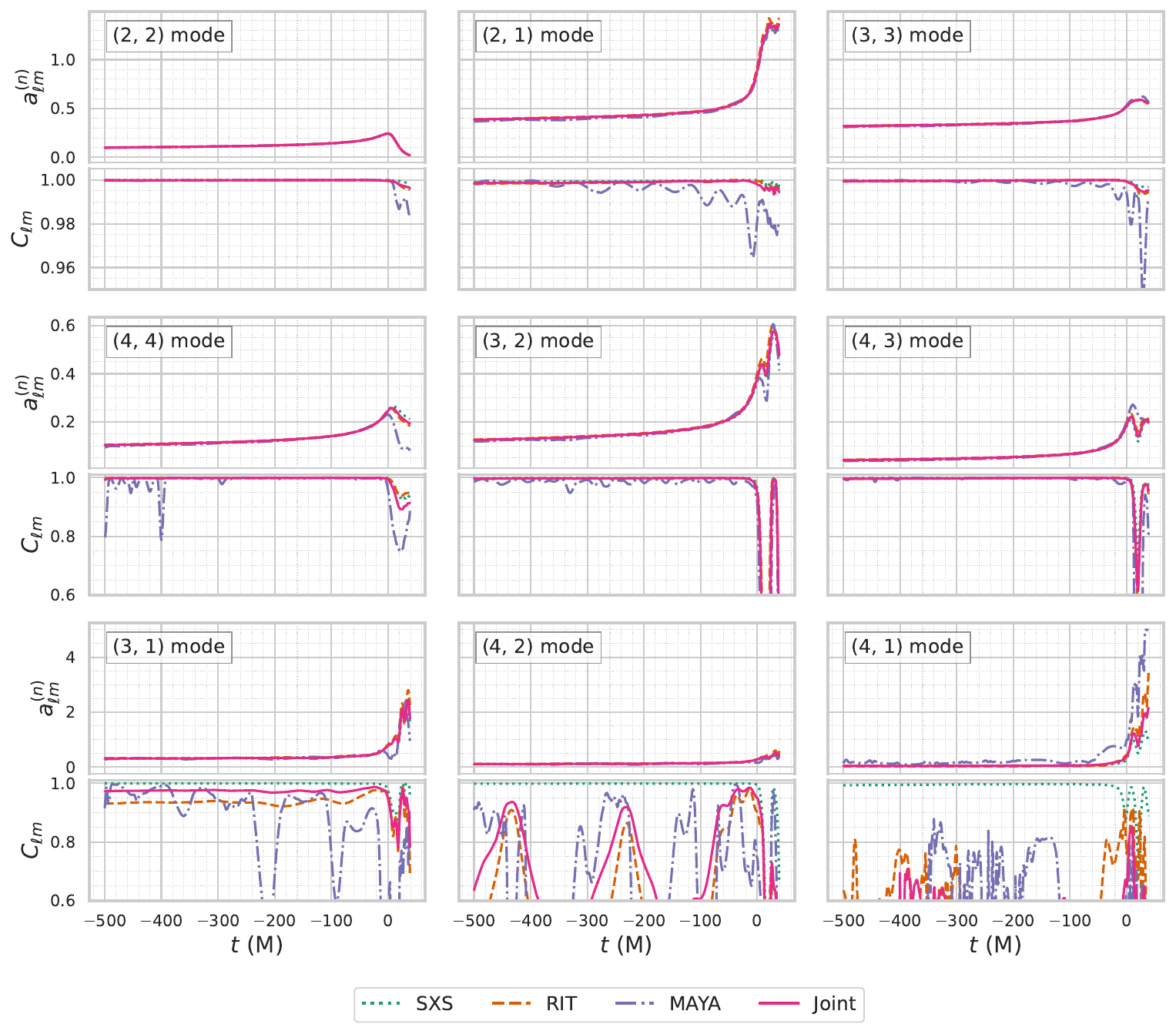}
    \caption{\label{fig:nonspinning_params} Leading-order fits for nonspinning simulations: Optimal fit coefficients $a_{\ell m}^{(n)}$ and Pearson correlation $C_{\ell m}$ for the leading-order fits defined in Eqs.~(\ref{eq:nonspinning_lo_fits}), shown as a function of time $t$ after the peak of the $(2,2)$ mode. The dotted, dashed, and dash-dotted curves show results obtained from the \ac{sxs}, \ac{rit}, and MAYA datasets, respectively. The solid curve shows results obtained from the joint dataset.}
\end{figure*}

Having established the relationship between $v$ and the optimal fit coefficients at early times, Fig.~\ref{fig:nonspinning_params} shows the fit coefficients $a_{\ell m}^{(n)}$ obtained near the merger using individual catalogs and the joint dataset. The Pearson correlation coefficient is also shown as a function of time for each mode to quantify the agreement between the data and the fits.

For the all modes, the recovered fit coefficients are approximately consistent across catalogs and vary smoothly in time, suggesting the fits are stable. The Pearson correlation remains above $0.99$ at all times for the $(2, 2)$, $(2,1)$, and $(3,3)$ modes across catalogs, except briefly for the MAYA dataset, which contributed the fewest simulations to our dataset. 

For the weaker modes, there is significantly more variability across catalogs. The \ac{sxs} catalog, which contributed the most simulations to our full dataset, shows the most stable behavior. It is followed by the \ac{rit} dataset, which shows erratic behavior only in the Pearson correlation for the $(4, 2)$ and $(4, 1)$ modes, but otherwise performs similarly to the \ac{sxs} dataset. The MAYA dataset is stable for the $(3,2)$ and $(4, 3)$ modes, but unstable for the $(3, 1)$, $(4, 2)$, and $(4, 1)$ modes. These subdominant modes have lower amplitudes and are more susceptible to numerical errors. Additionally, higher-$\ell$ modes have higher frequencies and more complex sourcing, making them more susceptible to resolution-dependent differences \cite{Scheel:2025jct,lovelace2016modeling}. 

For all datasets, as the time approaches $t=0M$, the coefficients tend to grow as each mode absorbs increasingly significant higher-order corrections. Focusing on the results for the \ac{sxs} data, we find that the fits maintain high correlations and vary smoothly over time for all modes throughout most of their evolution, up to $t \approx -10M$. However, the fit quality drops sharply around $t = 0M$ for all weaker modes. This decline indicates that higher-order corrections introduce markedly different dependencies on $\eta$ that cannot be captured by the leading-order basis alone, and additional $\eta$ dependencies must be introduced. 

\subsubsection{Aligned-spin simulations}
To reduce the impact of cross-catalog differences, we perform fits on the aligned-spin simulations using only the \ac{sxs} dataset. Nonetheless, Appendix~\ref{app:alignedspin_snapshots} presents snapshots of the \ac{sxs} fits with \ac{rit} and MAYA data points included for visual comparison across catalogs.

\begin{figure*}
    \centering
    \includegraphics[width=\linewidth]{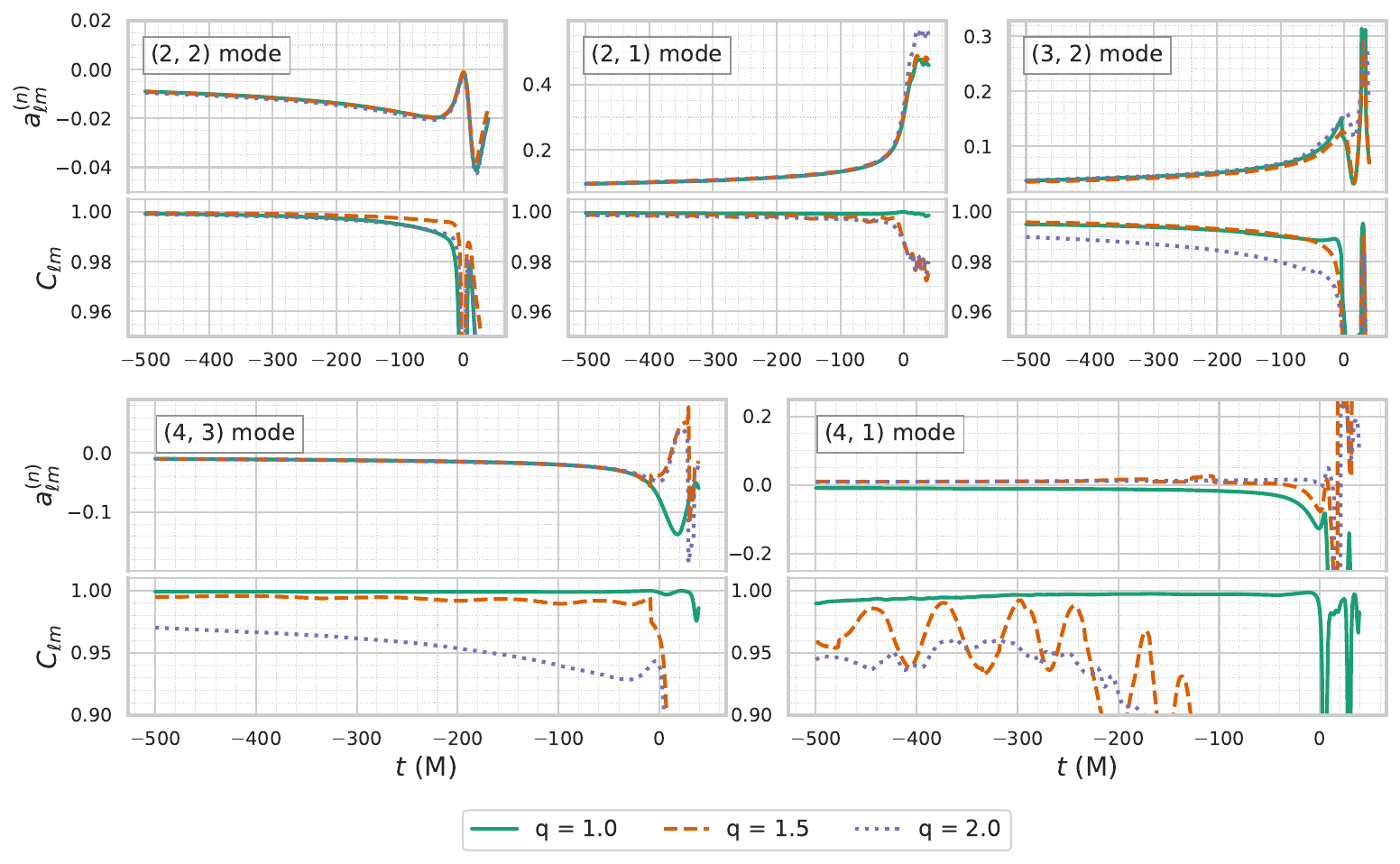}
    \caption{Leading-order fits for aligned-spin simulations: Optimal fit coefficients $a_{\ell m}^{(n)}$ and Pearson correlation $C_{\ell m}$ for the leading-order fits defined in Eqs.~(\ref{eq:alignedspin_lo_fits}), shown as a function of time $t$ after the peak of the $(2,2)$ mode. Solid, dashed, and dotted lines show results for simulations with $q=1.0$, $q=1.5$, and $q=2.0$, respectively. All fits were performed using simulations from the \ac{sxs} catalog only.}
    \label{fig:lower_order_fits_alignedspin}
\end{figure*}

Figure~\ref{fig:lower_order_fits_alignedspin} shows the fit parameters over time. Early on, recovered parameters vary little across mass ratios. This is because higher-order corrections depending solely on mass ratio are accounted for by the nonspinning \ac{nr} amplitudes $\hat{A}_{\ell m}^{(\mathrm{ns})}$ used in the fits. The consistency across mass ratios up to $t \approx -20M$ indicates that higher-order spin corrections introduce little to no additional mass-ratio dependence, beyond that captured at leading order. However, differences become more pronounced after the merger, where higher-order spin effects with distinct mass-ratio dependencies play a more significant role.

Regarding fit quality, the $(2,1)$ mode appears to be well described by its leading-order spin dependence through merger, with $C_{21}>0.99$ for $q=1.0$ and $C_{21}>0.97$ for $q=1.5$ and $q=2.0$ at all times. This is consistent with the results of Ref.~\cite{borhanian2020comparison}. Although the $(2, 2)$ and $(3, 2)$ modes maintain a high correlation with the fits until very close to merger, they deviate significantly from the \ac{pn} dependence near merger for all mass ratios. For the $(4, 3)$ mode, $C_{43} > 0.97$ for $q=1.0$ at all times, but this correlation decreases with increasing mass ratio, especially near merger.

Finally, the $(4, 1)$ fit also performs well for $q=1.0$ up to $t\approx0M$, but drops sharply near merger, where the coefficient suddenly increases in magnitude. Furthermore, for $q=1.5$ and $q=2.0$, the $(4, 1)$ fit is unstable throughout the evolution, showing small jumps in $a_{41}^{(4)}$ and fluctuations in $C_{41}$. This suggests the model is not well constrained by the data and is likely prone to overfitting the amplitudes for this mode. This uncertainty in the model is evident in the fit snapshots shown in Figs.~\ref{fig:alignedspin_q1.0}--\ref{fig:alignedspin_q2.0} in Appendix~\ref{app:alignedspin_snapshots}.

The result in \cite{borhanian2020comparison} found better agreement for higher mass ratios in the $(4, 1)$ mode. However, their analysis allowed a constant offset to vary freely, whereas we fixed this offset by enforcing the nonspinning limit through our $\hat{A}_{\ell m}^{(\mathrm{ns})}$. This suggests that some higher-order spin effects were possibly absorbed into their constant offset and may not be captured in our approach.

\subsection{Higher-order fits}\label{sec:results_ho}
We now investigate the extent to which higher-order dependencies are needed to significantly improve fit quality for nonspinning and aligned-spin simulations.

\subsubsection{Nonspinning simulations}
\begin{figure*}
\includegraphics[width=\linewidth]{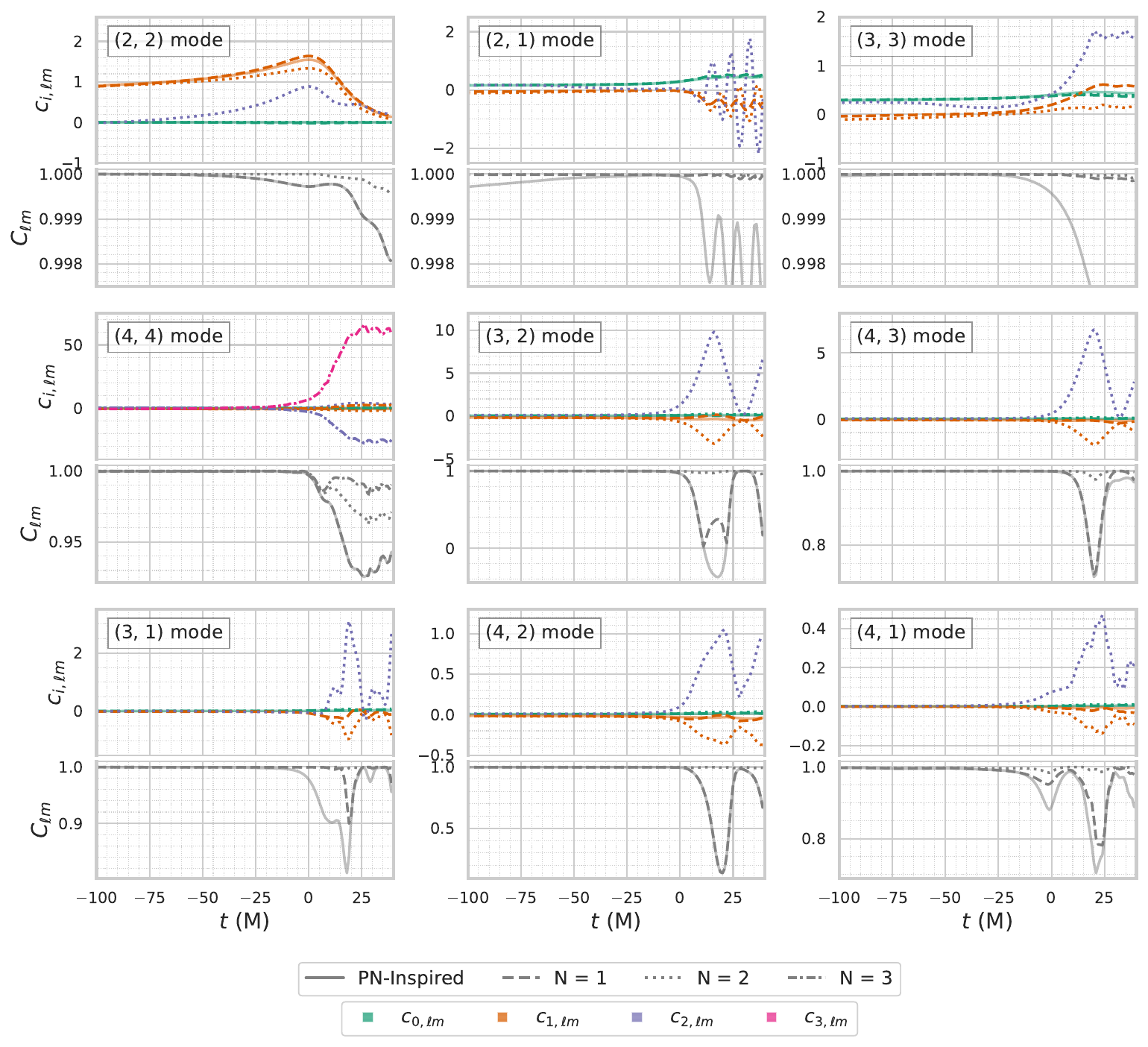}
    \caption{\label{fig:ho_nonspinning_params} Higher-order fits for nonspinning simulations: Optimal fit coefficients $c_i$ and Pearson correlation $C_{\ell m}$ for models with $N=1$ and $N=2$ as defined in Eq.~(\ref{eq:nonspinning_ho_fits}), shown as a function of time $t$ after the peak of the $(2,2)$ mode. Results for $N=3$ are only shown for the $(4, 4)$ mode. Solid curves show the fit coefficient obtained for the leading-order (\ac{pn}-inspired) fits in Eqs.~(\ref{eq:nonspinning_lo_fits}), converted to their corresponding $c_i$. Dashed, dotted, and dash-dotted curves show results for $N=1$, $N=2$, and $N=3$, respectively, with colors corresponding to different coefficients $c_i$.}
\end{figure*}

\begin{figure*}
\includegraphics[width=\linewidth]{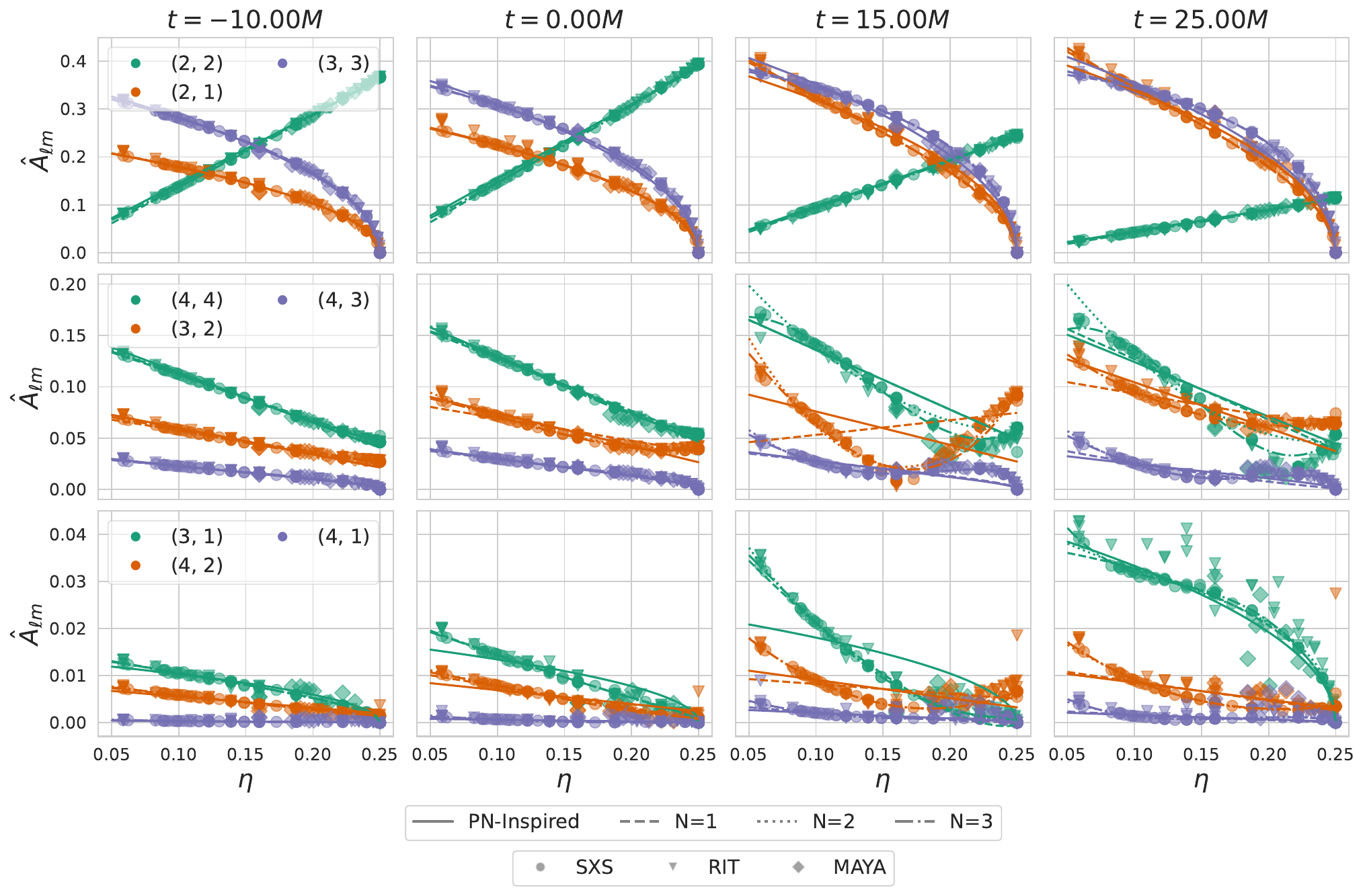}
    \caption{\label{fig:ho_nonspinning_snapshots} Higher-order fit snapshots for nonspinning simulations: Snapshots of the $(\ell, m)$ mode amplitudes, normalized to the $(2,2)$ mode, at selected times $t$ relative to the peak of the $(2, 2)$ mode, as a function of the symmetric mass ratio $\eta$. Line styles correspond to polynomial fits in $\eta$ at different orders $N$, as defined in Eq.~\eqref{eq:nonspinning_ho_fits}. Colors indicate the modes displayed in each row, and markers denote the \ac{nr} catalog associated with each data point. The best-fit curves shown were obtained using data from the \ac{sxs} catalog only.}
\end{figure*}

The impact of higher-order corrections on the quality of fits for nonspinning systems is illustrated in two complementary figures. Figure~\ref{fig:ho_nonspinning_params} shows the recovered fit coefficients and Pearson correlations when applying fits of degrees $N = 1$ and $N = 2$ to each mode, while Fig.~\ref{fig:ho_nonspinning_snapshots} provides snapshots of the amplitude data and these fits at selected times. Together, these figures allow for a visual assessment of how the inclusion of higher-order terms improves modeling accuracy across modes and times.

Figure~\ref{fig:ho_nonspinning_params} focuses on times $t \in [-100M, 40M]$, where higher-order $\eta$ terms are expected to have larger impact. At earlier times ($t \ll 0M$), leading-order fits were already found to perform well across all modes in the results presented in the previous section, as well as previous work in Ref.~\cite{borhanian2020comparison}. Furthermore, up to $t\sim -100M$, the higher-order fits recover coefficients consistent with leading-order fits, as can be seen in the figure.

As the system approaches merger ($t \to 0$), and especially in the postmerger regime ($t > 0$), higher-order $\eta$ dependencies become increasingly relevant, leading to noticeable improvements in fit quality for several modes. We find that for all modes except the $(4, 4)$ mode, the $N=3$ fit offers minimal improvement over the $N=2$ fit. Furthermore, the recovered parameters for the $N=3$ model vary nonsmoothly in time and increase to large values, indicating possible overfitting. We therefore only show fits for $N=1$ and $N=2$ in Fig.~\ref{fig:ho_nonspinning_params}, except for the $(4, 4)$ mode, which we discuss below. We observe the following trends:
\begin{enumerate}
    \item $(2, 2)$ mode: As previously found, the leading-order \ac{pn}-inspired fit remains well behaved through the merger. Adding a constant offset ($N=1$) does not change the fit appreciably, while adding an $\eta^2$ dependence slightly improves the correlation for $t\gtrsim-25M$.
    \item $(2,1)$ and $(3,3)$ modes: These modes slightly benefit from the inclusion of a $\delta\eta$ term ($N=1$). Improvements are noticeable after the merger, as seen at $t=15M$ and $t=25M$ in Fig.~\ref{fig:ho_nonspinning_snapshots}. However, improvements beyond $N=1$ are minimal.
    \item $(4,4)$ mode: Higher-order terms become significantly more important after the merger. In this regime, the $N=2$ and $N=3$ fits yield larger $C_{41}$ values than the lower-order models, and this improvement is clearly visible in Fig.~\ref{fig:ho_nonspinning_snapshots}. For the $N=3$ model, we find that the fit coefficients  $c_2$ and $c_3$ saturate the bounds $[-10, 10]$ near merger, although they vary smoothly and remain well within these bounds at earlier times. To ensure that the fit is not constrained by the imposed bounds, we extend the allowed range to $[-100, 100]$ for $t>0M$. While this extension could raise concerns about overfitting, we explicitly test this possibility in Appendix~\ref{app:bayesian_44} by performing a full Bayesian linear regression using the $N=3$ model. The Bayesian results are consistent with those obtained using differential evolution, and the preference for an $\eta^3$ dependence is supported by both methods.
    \item Subdominant modes with $\ell = 3,4$ and $m < \ell$: These modes, previously identified as poorly described by leading-order \ac{pn} fits near merger, show a strong preference for higher-order fits near and beyond the merger. Nevertheless, the $N=1$ and $N=2$ fits can capture their amplitude evolution with correlations comparable to the fits for the stronger modes. These improvements are apparent at late times in the snapshots in Fig.~\ref{fig:ho_nonspinning_snapshots}.
\end{enumerate}

Figure~\ref{fig:ho_nonspinning_snapshots} also plots data points from the \ac{rit} and MAYA catalogs. Although these were not used to generate the fits, they generally align well with the \ac{sxs}-based fits. However, cross-catalog dispersion increases at later times for subdominant modes, particularly for the $(4, 2)$ and $(4, 1)$ modes. 

\subsubsection{Aligned-spin simulations}

\begin{figure*}
\includegraphics[width=\linewidth]{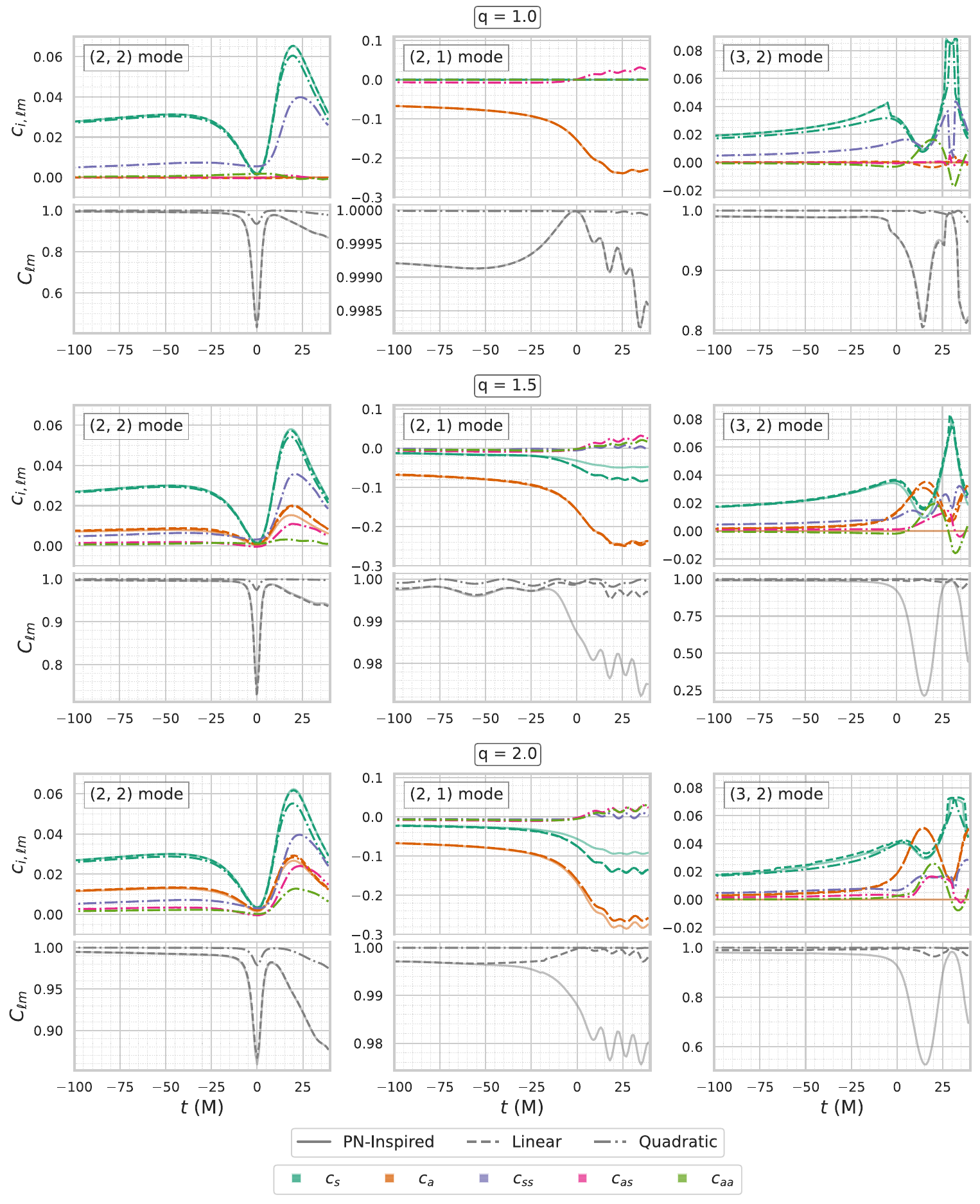}
    \caption{\label{fig:ho_alignedspin_params} Higher-order fits for aligned-spin simulations: Optimal fit coefficients $c_{i, \ell m}^{(n)}$ and Pearson correlation $C_{\ell m}$ for the linear and quadratic fits defined in Eqs. \ref{eq:alignedspin_ho_fits}. The top, middle, and bottom rows show fits for $q=1.0$, $q=1.5$, $q=2.0$ systems, respectively. Fits were performed using simulations from the \ac{sxs} catalog only.}
\end{figure*}

Figure~\ref{fig:ho_alignedspin_params} plots the recovered fit coefficients and Pearson correlations of the higher-order aligned-spin fits $\hat{A}_{lm}^{(i, \text{S})}$ defined in Eqs.~(\ref{eq:alignedspin_ho_fits}) for the $(2, 2)$, $(2, 1)$, and $(3, 2)$ modes, with the corresponding values from the leading-order fits included for reference. To further illustrate the behavior of the fits, Figs.~\ref{fig:alignedspin_q1.0}--\ref{fig:alignedspin_q2.0} present snapshots of the amplitude data and their associated fits at selected times for each mass ratio considered.

The $c_s$ and $c_a$ for the linear and quadratic models closely follow their corresponding values from the leading-order fits up to near merger, as expected given the good performance of the leading-order fits at earlier times. For the $(2, 2)$ mode, the $\chi_s^2$ contributions increase significantly after merger, and the quadratic fit yields a much higher Pearson correlation than both the linear and leading-order fits. For $q = 1.0$, the coefficients $c_a$, $c_{as}$ and $c_{aa}$ remain small for the $(2, 2)$ mode, but increase with increasing mass ratios. Similarly, the spin dependence of the $(3, 2)$ mode is better captured by the quadratic fit than the lower-order models. In this case, contributions from $\chi_s^2$ and $\chi_{aa}$ are significant around merger for equal-mass systems, while contributions from $\chi_a$ and $\chi_{a}\chi_s$ increase with increasing mass ratio. The improvement of the quadratic fit over the leading-order fit for the $(2, 2)$ and $(3, 2)$ modes is visually evident in Figs.~\ref{fig:alignedspin_q1.0}--\ref{fig:alignedspin_q2.0}. For the $(2, 1)$ mode, the leading-order fits already yield high correlations. Nevertheless, the added flexibility of the linear fits increases the correlation for unequal-mass systems near and after merger. Quadratic fits provide further improvement, although, as seen in Figs.~\ref{fig:alignedspin_q1.0}--\ref{fig:alignedspin_q2.0}, this improvement is modest.

For the $(4, 3)$ and $(4, 1)$ modes, we find that the fits are sensitive to the stochastic nature of differential evolution, particularly at higher mass ratios. This sensitivity likely arises from degeneracies among the five fit coefficients combined with relatively sparse coverage of simulations. Rather than using global optimization, we therefore perform Bayesian linear regression for these modes, as outlined in Appendix~\ref{app:bayesian}. Figure~\ref{fig:bayesian_43_41} shows the posterior distributions of each parameter as a function of time obtained using the quadratic fit. For equal-mass systems, the parameters of the $(4, 3)$ mode are well constrained and consistent with those of the leading-order fit. The inclusion of the $c_{as}$ term slightly improves the Pearson correlation relative to the leading-order fit after the merger. In contrast, for the $(4, 1)$ mode in equal-mass systems, in equal-mass systems, the fit exhibits occasional spikes, and improvement over the leading-order fit is not clearly established.

For $q=1.5$ systems, the simulations used do not sufficiently constrain the parameters of either the $(4, 3)$ or $(4, 1)$ modes. This is likely due to the smaller number of data points at this mass ratio, combined with the increasing importance of higher-order spin dependencies. As illustrated by the $q=2.0$ case, the parameters are better constrained when more simulations are available. However, higher-order spin effects become significant near merger, and the model fails to capture a clean evolution of the coefficients in this regime. These results suggest that higher-order spin dependencies are particularly important around merger for these modes, and that additional data would be necessary to robustly constrain the evolution of their coefficients.

\begin{figure*}
    \centering
    \includegraphics[width=\linewidth]{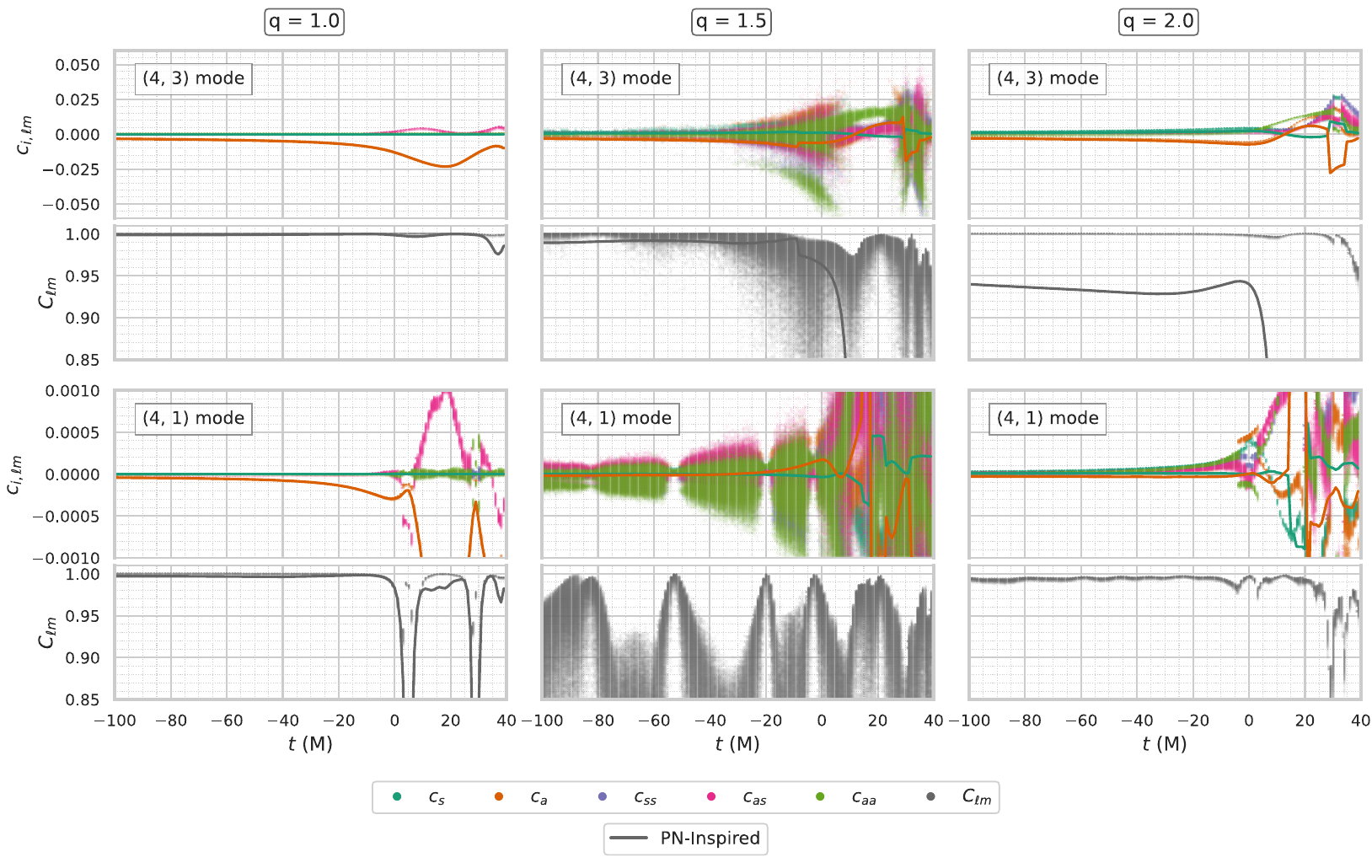}
    \caption{Posterior distributions for the $c_i$ of the quadratic model in Eq.~(\ref{eq:alignedspin_ho_quadratic}) over time for the $(4, 3)$ and $(4, 1)$ modes.}
    \label{fig:bayesian_43_41}
\end{figure*}

Overall, the inclusion of quadratic spin terms improves the modeling of the $(2, 2)$ and $(3, 2)$ modes for all mass ratios considered and of the $(4, 3)$ mode for mass ratios $q>1$ after the merger. Visual comparison of the \ac{rit} and MAYA simulations in Figs.~\ref{fig:alignedspin_q1.0}--\ref{fig:alignedspin_q2.0} confirms cross-catalog consistency with these improved higher-order fits for the $(2, 2)$, $(2, 1)$, $(3, 2)$, and $(4, 3)$ modes. For the $(4,1)$ mode, however, there is greater dispersion across catalogs, again likely due to differences in numerical techniques across catalogs.

\section{\label{sec:conclusions}Conclusions}
In this study, we analyzed the mode amplitudes of quasicircular, nonprecessing \ac{bbh} mergers using \ac{nr} data. Our primary goal was to understand the extent to which the \ac{pn} dependence on mass ratios and spins is carried over from the inspiral, particularly near and after the merger, where modeling remains challenging.

We obtained fit coefficients for the mode amplitudes relative to the $(2, 2)$ mode as a function of time and verified their stability across the waveform, from late inspiral ($t=-500M$) to postmerger ($t=+40M$). The stability and quality of these fits demonstrate that for several modes, the time dependence can be cleanly separated from the dependence on the mass ratio and spin. This separation provides a framework for constructing simplified waveform models in the near- and postmerger regime by directly modeling the time evolution of the fit coefficients.

In the inspiral regime, we connected the fit coefficients to the \ac{pn} velocity. As the system evolves toward the merger, the coefficients deviate from the \ac{pn} predictions, but we interpret these deviations as absorptions of higher-order effects that preserve the same overall dependence on mass ratio and spin. This reinterpretation provides intuition behind why these \ac{pn}-inspired fits can, for some modes, be applied in the strong-field regime.

To explore higher-order parameter dependencies, we tested fits of increasing complexity in both mass ratio and spin. For nonspinning systems, we found that \ac{pn}-inspired fits with polynomial order $N \leq 3$ can model mode amplitudes accurately even close to the merger. While leading-order fits remain effective through the late inspiral, selectively incorporating higher-order terms substantially improves the modeling of subdominant modes near and after the merger. For aligned-spin simulations, we found that quadratic fits in $\chi_s$ and $\chi_a$ offer notable improvements for modes that were not well modeled by the leading-order \ac{pn} fits, such as the $(2, 2)$ and $(3,2)$ modes. However, the large number of free parameters in quadratic fits can lead to instabilities. For higher-order modes of unequal-mass systems, for which there are fewer available simulations and higher-order spin dependencies become more important, these models are not well constrained.

We validated these results across the \ac{sxs}, \ac{rit}, and MAYA catalogs. While most fits were performed using only \ac{sxs} data, we generally found good visual agreement with the other catalogs. Discrepancies in a small number of cases may point to numerical artifacts and can serve as indicators of outlying simulations.

The \ac{pn} expansion does not capture physical effects that are unique to the strong-field regime. For example, near merger, the deformation of the black holes’ dynamical horizons leads to tidal heating through the flow of energy and angular momentum through the horizons. These processes increase the horizon area and change the masses and spins of the black holes, effects that are not typically incorporated in \ac{pn} waveforms \cite{Prasad:2024vsz, Chia:2020yla, Chatziioannou:2012gq, Price:2001un}. We therefore emphasize that the phenomenological fits presented here are merely inspired by the functional forms of leading-order PN mode amplitudes. Their physical interpretation cannot be fully described within a PN framework, as they effectively absorb contributions from strong-field, near-merger physics that lie beyond the formal domain of validity of the PN approximation.

Our conclusions do not address the accuracy of \ac{pn}-inspired fits for modeling the phase evolution. A complementary analysis of phase, which would require additional methodological considerations, is left for future work. We also aim to extend this analysis to simulations of precessing systems, in which the spin components are time dependent. 

Overall, this study advances the understanding of mode amplitude structure in \ac{bbh} mergers and lays a framework for efficient amplitude-consistent waveform modeling, particularly near the merger.

\begin{acknowledgments}
We wish to acknowledge B. S. Sathyaprakash for his invaluable guidance and insightful comments throughout this work. We are grateful to Vaishak Prasad for his assistance with the \ac{sxs} catalog and for providing useful comments on this manuscript. We further acknowledge the \ac{sxs}, \ac{rit}, and MAYA Collaborations for their publicly available catalogs of \ac{nr} simulations, which were used in this study. We also thank Carlos Lousto for guidance in using the \ac{rit} catalog, and Deborah Ferguson and Aasim Jan for their help with \textsc{mayawaves} and the MAYA catalog. Finally, we are grateful to Ssohrab Borhanian for answering questions related to his original study and to Koustav Chandra, Rossella Gamba, and Dongze Sun for helpful discussions.
\end{acknowledgments}

\appendix
\section{\label{app:bayesian}Bayesian linear regression}
When performing Bayesian regression, we assume the NR amplitudes $\hat{A}_{\ell m}^{(\mathrm{NR})}$ are described by one of the models in Eqs.~\eqref{eq:alignedspin_lo_fits}-\eqref{eq:alignedspin_ho_fits}, which are characterized by a set of fit coefficients $\{ c_i \} \equiv \vec{c}$. We use a Gaussian likelihood with unknown variance $\mathcal{L}( \hat{A}_{\ell m}^{(\mathrm{NR})}|\, \vec{c}\, )$ and uniform priors $\pi(c_i) = \mathcal{U}(c_{min}, c_{max})$ to compute the posterior distribution for each $c_i$ from Bayes’s theorem,
\begin{align}
    p( \, \vec{c} \, | \hat{A}_{\ell m}^{(\mathrm{NR})}) = \frac{\mathcal{L}( \hat{A}_{\ell m}^{(\mathrm{NR})}|\, \vec{c}\, ) \pi (\vec{c} )}{\mathcal{Z}( \hat{A}_{\ell m}^{(\mathrm{NR})} )},
\end{align}
where $\mathcal{Z}( \hat{A}_{\ell m}^{(\mathrm{NR})} )$ is the normalizing evidence,
\begin{align}
    \mathcal{Z}( \hat{A}_{\ell m}^{(\mathrm{NR})} ) = \int \mathcal{L}( \hat{A}_{\ell m}^{(\mathrm{NR})}|\, \vec{c}\, ) \pi (\vec{c} ) \, d\vec{c}
\end{align}
which is especially useful for model comparison. In particular, the log Bayes factor between models $1$ and $2$,
\begin{align}
    \log BF_{12} = \log \mathcal{Z}_1 - \log \mathcal{Z}_2
\end{align}
is positive if model $1$ is preferred and negative when model $2$ is preferred.

We compute posteriors using the Dynesty nested-sampling algorithm \cite{Speagle:2019ivv} accessed through the \textsc{bilby} package \cite{Ashton:2018jfp,Romero-Shaw:2020owr}. 

\subsection{\label{app:bayesian_44}$N=3$ fit for the $(4, 4)$ mode}
As discussed in Sec.~\ref{sec:results_ho}, the coefficients $c_2$ and $c_3$ of the $N=3$ model rail against the previously imposed bounds on the parameters $[-10, 10]$ for the $(4, 4)$ mode. To examine whether extending the prior bounds on $c_2$ and $c_3$ leads to overfitting the data, we perform Bayesian regression using uniform priors $\pi(c_i) = \mathcal{U}(-100, 100)$ for all $\{c_i\}$.

\begin{figure*}
\includegraphics[width=\linewidth]{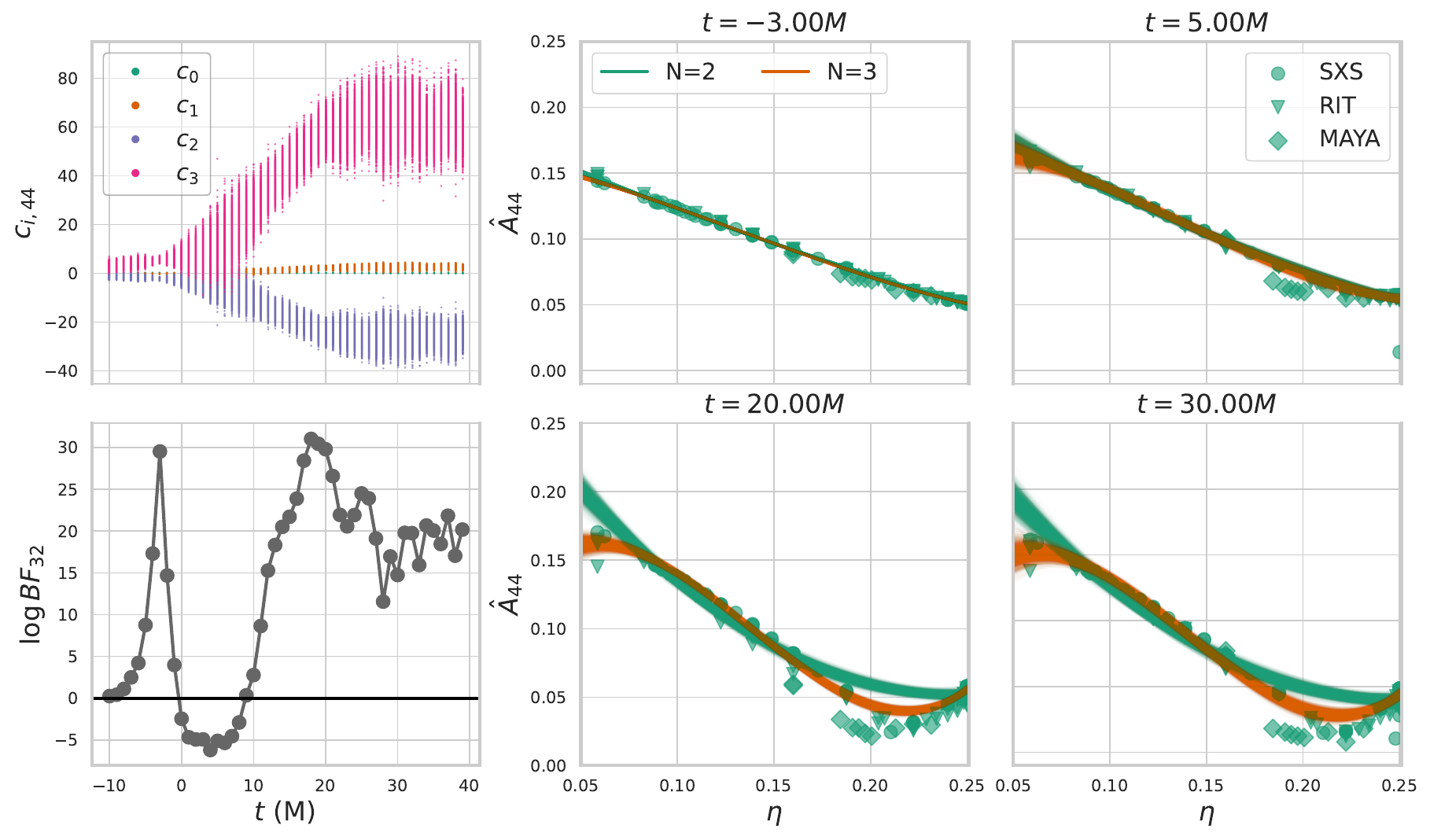}
    \caption{\label{fig:bayesian_44} Results of Bayesian linear regression for the $(4, 4)$ mode amplitudes for nonspinning systems. The top-left panel shows the posterior distributions of each coefficient $c_i$ as a function of time. The bottom-left panel shows the log Bayes factor for the $N=3$ model relative to the $N=2$ model. The middle and right columns show snapshots of the fits at selected times, with realizations drawn from the posterior distributions for the $N=2$ (teal) and $N=3$ (orange) models.}
\end{figure*}

Figure~\ref{fig:bayesian_44} shows the posterior distributions obtained at each time step, together with snapshots of the fitted models constructed from posterior samples. All coefficients remain well within the prior range and are consistent with the results presented in Sec.~\ref{sec:results_ho} obtained using the differential evolution algorithm. The coefficients $c_2$ and $c_3$ exhibit relatively large uncertainties. As shown in the snapshots, this uncertainty mostly affects predictions at small $\eta$ and is likely driven by the limited number of data points in this regime, i.e., the limited number of simulations for systems with high-mass ratio. While the data do not tightly constrain the exact value of $c_3$, they favor the presence of the $\eta^3$ term over its absence. This is evidenced by the positive value of the log Bayes factor at late times and visually corroborated in the snapshots.

\section{Selected simulations} \label{app:simulations_info}
Tables~\ref{tab:nonspinning_all} and \ref{tab:alignedspin_all} list the simulations from the \ac{sxs}, \ac{rit}, and MAYA catalogs that were used in this study, following the selection criteria outlined in Sec.~\ref{sec:methods_NR_catalogs}. Table~\ref{tab:nonspinning_all} lists the nonspinning simulations with their mass ratio, while Table~\ref{tab:alignedspin_all} lists the aligned-spin simulations with their mass ratio and dimensionless spin components.

\begin{table*}
\caption{\label{tab:nonspinning_all} List of 103 nonspinning simulations from the \ac{sxs}, \ac{rit}, and MAYA catalogs used in this work, grouped by mass ratio $q$.}
\begin{ruledtabular}
\begin{tabular}{ccccccccc}
$q$ & \multicolumn{2}{c}{ID(s)} & $q$ & \multicolumn{2}{c}{ID(s)} & $q$ & \multicolumn{2}{c}{ID(s)} \\ 
\cline{1-9}
$1.00$ & \texttt{SXS:BBH:0389} & \texttt{SXS:BBH:2377} & $2.60$ & \texttt{GT0651} &    & $7.00$ & \texttt{SXS:BBH:2491} & \texttt{RIT:BBH:0416} \\ 
     & \texttt{SXS:BBH:2378} & \texttt{SXS:BBH:3624} &    &    &     &    &    &     \\ 
     & \texttt{SXS:BBH:4434} & \texttt{SXS:BBH:2496} &    &    &     &    &    &     \\ 
     & \texttt{SXS:BBH:1132} & \texttt{SXS:BBH:3864} &    &    &     &    &    &     \\ 
     & \texttt{SXS:BBH:1154} & \texttt{SXS:BBH:1155} &    &    &     &    &    &     \\ 
     & \texttt{SXS:BBH:1153} & \texttt{SXS:BBH:3617} &    &    &     &    &    &     \\ 
     & \texttt{SXS:BBH:2375} & \texttt{SXS:BBH:3634} &    &    &     &    &    &     \\ 
     & \texttt{SXS:BBH:2376} & \texttt{SXS:BBH:2325} &    &    &     &    &    &     \\ 
     & \texttt{SXS:BBH:2326} & \texttt{RIT:BBH:0112} &    &    &     &    &    &     \\ 
     & \texttt{RIT:eBBH:1090} &    &    &    &     &    &    &     \\ 
\hline 
$1.08$ & \texttt{RIT:BBH:0458} &    & $2.70$ & \texttt{GT0650} &    & $7.19$ & \texttt{SXS:BBH:0188} &    \\ 
\hline 
$1.18$ & \texttt{RIT:BBH:0084} & \texttt{RIT:BBH:0113} & $2.80$ & \texttt{GT0653} &    & $7.50$ & \texttt{SXS:BBH:2490} &    \\ 
\hline 
$1.20$ & \texttt{SXS:BBH:0198} &    & $2.90$ & \texttt{GT0654} &    & $7.76$ & \texttt{SXS:BBH:0195} &    \\ 
\hline 
$1.33$ & \texttt{RIT:BBH:0045} & \texttt{RIT:BBH:0114} & $3.00$ & \texttt{SXS:BBH:1179} & \texttt{SXS:BBH:2498} & $8.00$ & \texttt{SXS:BBH:2707} & \texttt{SXS:BBH:3622} \\ 
     & \texttt{RIT:eBBH:1241} &    &      & \texttt{SXS:BBH:2265} & \texttt{RIT:BBH:0102} &    &    &     \\ 
    &    &    &      & \texttt{RIT:BBH:0138} &     &    &    &     \\ 
\hline 
$1.50$ & \texttt{SXS:BBH:2331} & \texttt{SXS:BBH:3984} & $3.10$ & \texttt{GT0758} &    & $8.27$ & \texttt{SXS:BBH:0186} &    \\ 
     & \texttt{SXS:BBH:0593} & \texttt{RIT:BBH:0085} &    &    &     &    &    &     \\ 
     & \texttt{RIT:BBH:0115} &    &    &    &     &    &    &     \\ 
\hline 
$1.67$ & \texttt{RIT:BBH:0086} & \texttt{RIT:BBH:0116} & $3.50$ & \texttt{SXS:BBH:2483} &    & $8.73$ & \texttt{SXS:BBH:0199} &    \\ 
\hline 
$1.75$ & \texttt{GT0727} &    & $4.00$ & \texttt{SXS:BBH:2485} & \texttt{SXS:BBH:3631} & $9.00$ & \texttt{SXS:BBH:2495} &    \\ 
    &    &    &      & \texttt{SXS:BBH:2499} & \texttt{SXS:BBH:1220}  &    &    &     \\ 
    &    &    &      & \texttt{RIT:BBH:0088} & \texttt{RIT:BBH:0119}  &    &    &     \\ 
    &    &    &      & \texttt{RIT:eBBH:1133} & \texttt{GT0454}  &    &    &     \\ 
    &    &    &      & \texttt{MAYA1040} &     &    &    &     \\ 
\hline 
$1.82$ & \texttt{RIT:BBH:1020} &    & $4.50$ & \texttt{SXS:BBH:2484} & \texttt{SXS:BBH:3144} & $9.17$ & \texttt{SXS:BBH:0189} &    \\ 
\hline 
$2.00$ & \texttt{SXS:BBH:2497} & \texttt{SXS:BBH:1166} & $5.00$ & \texttt{SXS:BBH:2374} & \texttt{SXS:BBH:2487} & $9.20$ & \texttt{SXS:BBH:1108} &    \\ 
     & \texttt{SXS:BBH:1167} & \texttt{RIT:BBH:0077} &      & \texttt{SXS:BBH:3619} & \texttt{RIT:BBH:0089} &    &    &     \\ 
     & \texttt{RIT:BBH:0117} & \texttt{RIT:eBBH:1200} &      & \texttt{RIT:BBH:0120} & \texttt{RIT:BBH:0152} &    &    &     \\ 
     & \texttt{GT0446} &    &    &    &     &    &    &     \\ 
\hline 
$2.25$ & \texttt{GT0757} &    & $5.50$ & \texttt{SXS:BBH:2486} &    & $9.99$ & \texttt{SXS:BBH:0185} &    \\ 
\hline 
$2.32$ & \texttt{SXS:BBH:0201} &    & $6.00$ & \texttt{SXS:BBH:2164} & \texttt{SXS:BBH:2489} & $10.00$ & \texttt{RIT:BBH:0978} &    \\ 
    &    &    &      & \texttt{SXS:BBH:3630} & \texttt{RIT:BBH:0090}  &    &    &     \\ 
    &    &    &      & \texttt{RIT:BBH:0121} &     &    &    &     \\ 
\hline 
$2.41$ & \texttt{RIT:BBH:0139} &    & $6.50$ & \texttt{SXS:BBH:2488} &    & $14.00$ & \texttt{SXS:BBH:2480} &    \\ 
\hline 
$2.50$ & \texttt{RIT:BBH:0087} & \texttt{RIT:BBH:0118} & $6.58$ & \texttt{SXS:BBH:0192} &    & $15.00$ & \texttt{SXS:BBH:2477} & \texttt{RIT:BBH:0373} \\
    &    &    &    &    &    &      & \texttt{RIT:BBH:0942} & \texttt{RIT:BBH:0943} \\
    &    &    &    &    &    &      & \texttt{RIT:BBH:0957} &    
\end{tabular}
\end{ruledtabular}
\end{table*}

\begin{table*}
\caption{\label{tab:alignedspin_all} List of 172 aligned-spin simulations from the \ac{sxs}, \ac{rit}, and MAYA catalogs used in this work. Only \ac{sxs} simulations were used to construct the fits in Figs.\ref{fig:alignedspin_q1.0}--\ref{fig:alignedspin_q2.0}, but RIT and MAYA simulations are included for visual comparison.}
\begin{ruledtabular}
\begin{tabular}{cccc|cccc|cccc}
$q$ & ID & $\chi_{1z}$ & $\chi_{2z}$ &
$q$ & ID & $\chi_{1z}$ & $\chi_{2z}$ &
$q$ & ID & $\chi_{1z}$ & $\chi_{2z}$ \\
\hline
1.0 & \texttt{SXS:BBH:2100} & 0.5 & -0.9 & 1.0 & \texttt{SXS:BBH:1497} & 0.68 & 0.67 & 2.0 & \texttt{SXS:BBH:2131} & 0.85 & 0.85 \\
1.0 & \texttt{SXS:BBH:2093} & -0.5 & 0.9 & 1.0 & \texttt{SXS:BBH:3926} & 0.46 & -0.32 & 2.0 & \texttt{SXS:BBH:2132} & 0.87 & -0.85 \\
1.0 & \texttt{SXS:BBH:1492} & -0.47 & -0.79 & 1.0 & \texttt{SXS:BBH:0330} & -0.8 & 0.8 & 2.0 & \texttt{SXS:BBH:2109} & -0.6 & -0.6 \\
1.0 & \texttt{SXS:BBH:2099} & 0.4 & 0.8 & 1.0 & \texttt{SXS:BBH:2085} & -0.9 & 0.9 & 2.0 & \texttt{SXS:BBH:2127} & 0.5 & 0.5 \\
1.0 & \texttt{SXS:BBH:2098} & 0.4 & -0.8 & 1.0 & \texttt{SXS:BBH:2091} & -0.6 & 0.6 & 2.0 & \texttt{SXS:BBH:2114} & -0.3 & -0.3 \\
1.0 & \texttt{SXS:BBH:2095} & -0.4 & 0.8 & 1.0 & \texttt{SXS:BBH:1476} & -0.8 & 0.8 & 2.0 & \texttt{SXS:BBH:2116} & -0.3 & 0.3 \\
1.0 & \texttt{SXS:BBH:2094} & -0.4 & -0.8 & 1.0 & \texttt{SXS:BBH:2087} & -0.8 & 0.8 & 2.0 & \texttt{SXS:BBH:2123} & 0.3 & -0.3 \\
1.0 & \texttt{SXS:BBH:3912} & -0.47 & -0.79 & 1.0 & \texttt{SXS:BBH:3927} & 0.51 & 0.29 & 2.0 & \texttt{SXS:BBH:0448} & 0.4 & -0.4 \\
1.0 & \texttt{SXS:BBH:3982} & -0.18 & 0.72 & 1.0 & \texttt{SXS:BBH:1507} & 0.51 & 0.29 & 2.0 & \texttt{SXS:BBH:2130} & 0.6 & 0.6 \\
1.0 & \texttt{SXS:BBH:3923} & 0.14 & 0.73 & 1.0 & \texttt{SXS:BBH:0376} & 0.6 & -0.4 & 2.0 & \texttt{SXS:BBH:0554} & 0.2 & -0.4 \\
1.0 & \texttt{SXS:BBH:1502} & -0.42 & 0.7 & 1.0 & \texttt{SXS:BBH:0394} & 0.6 & 0.4 & 2.0 & \texttt{SXS:BBH:2111} & -0.6 & 0.6 \\
1.0 & \texttt{SXS:BBH:3922} & -0.42 & 0.7 & 1.0 & \texttt{SXS:BBH:0462} & -0.6 & -0.4 & 2.0 & \texttt{SXS:BBH:2112} & -0.5 & -0.5 \\
1.0 & \texttt{SXS:BBH:4066} & -0.95 & 0.95 & 1.0 & \texttt{SXS:BBH:0447} & -0.6 & 0.4 & 2.0 & \texttt{SXS:BBH:2510} & 0.6 & 0.6 \\
1.0 & \texttt{SXS:BBH:3977} & 0.95 & -0.95 & 1.0 & \texttt{SXS:BBH:2088} & -0.62 & -0.25 & 2.0 & \texttt{SXS:BBH:2506} & 0.5 & 0.5 \\
1.0 & \texttt{SXS:BBH:0436} & -0.2 & -0.4 & 1.0 & \texttt{SXS:BBH:2103} & 0.65 & 0.25 & 2.0 & \texttt{SXS:BBH:2501} & 0.25 & 0.25 \\
1.0 & \texttt{SXS:BBH:3534} & -0.8 & 0.8 & 1.0 & \texttt{SXS:BBH:3919} & -0.75 & 0.34 & 2.0 & \texttt{SXS:BBH:0513} & 0.6 & -0.4 \\
1.0 & \texttt{SXS:BBH:0304} & 0.5 & -0.5 & 1.0 & \texttt{SXS:BBH:3915} & 0.78 & 0.53 & 2.0 & \texttt{SXS:BBH:0387} & -0.6 & -0.4 \\
1.0 & \texttt{SXS:BBH:0459} & 0.2 & -0.4 & 1.0 & \texttt{SXS:BBH:3920} & -0.77 & -0.2 & 2.0 & \texttt{SXS:BBH:0614} & 0.75 & -0.5 \\
1.0 & \texttt{SXS:BBH:0370} & -0.2 & 0.4 & 1.0 & \texttt{SXS:BBH:1495} & 0.78 & 0.53 & 2.0 & \texttt{SXS:BBH:2125} & 0.3 & 0.3 \\
1.0 & \texttt{SXS:BBH:4029} & 0.95 & -0.95 & 1.0 & \texttt{SXS:BBH:3553} & 0.8 & 0.4 & 2.0 & \texttt{SXS:BBH:0412} & -0.2 & -0.4 \\
1.0 & \texttt{SXS:BBH:0327} & 0.8 & -0.8 & 1.0 & \texttt{SXS:BBH:3582} & -0.8 & -0.4 & 2.0 & \texttt{SXS:BBH:3981} & -0.12 & -0.65 \\
1.0 & \texttt{SXS:BBH:2092} & -0.5 & 0.5 & 1.0 & \texttt{SXS:BBH:3578} & -0.8 & 0.4 & 1.0 & \texttt{RIT:BBH:0016} & 0.8 & -0.8 \\
1.0 & \texttt{SXS:BBH:2422} & 0.6 & 0.6 & 1.0 & \texttt{SXS:BBH:1499} & -0.75 & 0.34 & 1.0 & \texttt{RIT:BBH:0228} & -0.4 & 0.85 \\
1.0 & \texttt{SXS:BBH:2419} & -0.2 & -0.2 & 1.0 & \texttt{SXS:BBH:3601} & 0.8 & -0.4 & 1.0 & \texttt{RIT:BBH:0261} & 0.4 & -0.85 \\
1.0 & \texttt{SXS:BBH:2418} & -0.44 & -0.44 & 1.0 & \texttt{SXS:BBH:1500} & -0.77 & -0.2 & 1.0 & \texttt{RIT:BBH:0324} & 0.9 & 0.9 \\
1.0 & \texttt{SXS:BBH:4072} & -0.95 & -0.95 & 1.0 & \texttt{SXS:BBH:2106} & 0.9 & 0.5 & 1.0 & \texttt{RIT:BBH:0499} & 0.95 & -0.95 \\
1.0 & \texttt{SXS:BBH:3629} & -0.9 & -0.9 & 1.0 & \texttt{SXS:BBH:2083} & -0.9 & -0.5 & 1.0 & \texttt{RIT:BBH:0573} & 0.95 & 0.0 \\
1.0 & \texttt{SXS:BBH:1475} & -0.8 & -0.8 & 1.5 & \texttt{SXS:BBH:0437} & -0.2 & -0.8 & 1.5 & \texttt{RIT:BBH:0289} & 0.8 & -0.8 \\
1.0 & \texttt{SXS:BBH:2102} & 0.6 & 0.6 & 1.5 & \texttt{SXS:BBH:0579} & 0.4 & -0.8 & 1.5 & \texttt{RIT:BBH:0616} & 0.95 & -0.95 \\
1.0 & \texttt{SXS:BBH:2500} & 0.25 & 0.25 & 1.5 & \texttt{SXS:BBH:2342} & -0.5 & 0.5 & 1.5 & \texttt{RIT:BBH:0687} & 0.95 & 0.95 \\
1.0 & \texttt{SXS:BBH:0160} & 0.9 & 0.9 & 1.5 & \texttt{SXS:BBH:0369} & 0.6 & -0.8 & 1.5 & \texttt{RIT:BBH:0788} & 0.95 & 0.0 \\
1.0 & \texttt{SXS:BBH:1122} & 0.44 & 0.44 & 1.5 & \texttt{SXS:BBH:0392} & -0.2 & 0.8 & 2.0 & \texttt{RIT:BBH:0083} & 0.5 & -0.5 \\
1.0 & \texttt{SXS:BBH:1123} & 0.5 & 0.5 & 1.5 & \texttt{SXS:BBH:2348} & 0.5 & -0.5 & 2.0 & \texttt{RIT:BBH:0245} & -0.85 & 0.5 \\
1.0 & \texttt{SXS:BBH:2505} & 0.5 & 0.5 & 1.5 & \texttt{SXS:BBH:0441} & 0.6 & 0.8 & 2.0 & \texttt{RIT:BBH:0247} & 0.85 & -0.5 \\
1.0 & \texttt{SXS:BBH:2089} & -0.6 & -0.6 & 1.5 & \texttt{SXS:BBH:0397} & -0.8 & -0.4 & 2.0 & \texttt{RIT:BBH:0296} & -0.85 & -0.5 \\
1.0 & \texttt{SXS:BBH:2512} & 0.7 & 0.7 & 1.5 & \texttt{SXS:BBH:1415} & 0.5 & 0.5 & 2.0 & \texttt{RIT:BBH:0314} & 0.85 & 0.5 \\
1.0 & \texttt{SXS:BBH:0178} & 0.99 & 0.99 & 1.5 & \texttt{SXS:BBH:0372} & 0.8 & -0.4 & 2.0 & \texttt{RIT:BBH:0336} & 0.0 & -0.8 \\
1.0 & \texttt{SXS:BBH:3897} & 0.8 & 0.8 & 2.0 & \texttt{SXS:BBH:0619} & 0.9 & 0.9 & 2.0 & \texttt{RIT:BBH:0338} & 0.0 & 0.8 \\
1.0 & \texttt{SXS:BBH:0176} & 0.96 & 0.96 & 2.0 & \texttt{SXS:BBH:2113} & -0.37 & 0.85 & 2.0 & \texttt{RIT:BBH:0446} & 0.5 & -0.8 \\
1.0 & \texttt{SXS:BBH:3895} & -0.8 & -0.8 & 2.0 & \texttt{SXS:BBH:2126} & 0.37 & -0.85 & 2.0 & \texttt{RIT:BBH:0447} & -0.25 & -0.25 \\
1.0 & \texttt{SXS:BBH:3627} & 0.95 & 0.95 & 2.0 & \texttt{SXS:BBH:2122} & 0.13 & 0.85 & 2.0 & \texttt{RIT:BBH:0451} & -0.5 & 0.8 \\
1.0 & \texttt{SXS:BBH:2086} & -0.8 & -0.8 & 2.0 & \texttt{SXS:BBH:0617} & 0.5 & 0.75 & 2.0 & \texttt{RIT:BBH:0452} & 0.25 & -0.25 \\
1.0 & \texttt{SXS:BBH:2509} & 0.6 & 0.6 & 2.0 & \texttt{SXS:BBH:2117} & -0.13 & -0.85 & 2.0 & \texttt{RIT:BBH:0454} & -0.25 & 0.25 \\
1.0 & \texttt{SXS:BBH:3628} & 0.97 & 0.97 & 2.0 & \texttt{SXS:BBH:0399} & 0.2 & 0.4 & 2.0 & \texttt{RIT:BBH:0457} & 0.25 & 0.25 \\
1.0 & \texttt{SXS:BBH:3518} & -0.8 & -0.8 & 2.0 & \texttt{SXS:BBH:0333} & 0.8 & 0.8 & 2.0 & \texttt{RIT:BBH:0462} & 0.85 & -0.25 \\
1.0 & \texttt{SXS:BBH:3625} & -0.8 & -0.8 & 2.0 & \texttt{SXS:BBH:0618} & 0.8 & 0.8 & 2.0 & \texttt{RIT:BBH:0465} & -0.85 & -0.25 \\
1.0 & \texttt{SXS:BBH:3976} & -0.95 & -0.95 & 2.0 & \texttt{SXS:BBH:4073} & 0.9 & 0.9 & 2.0 & \texttt{RIT:BBH:0466} & -0.85 & 0.25 \\
1.0 & \texttt{SXS:BBH:2420} & 0.2 & 0.2 & 2.0 & \texttt{SXS:BBH:4078} & 0.9 & -0.9 & 2.0 & \texttt{RIT:BBH:0477} & 0.85 & 0.25 \\
1.0 & \texttt{SXS:BBH:2421} & -0.6 & -0.6 & 2.0 & \texttt{SXS:BBH:4120} & -0.9 & -0.9 & 2.0 & \texttt{RIT:BBH:0679} & 0.95 & -0.95 \\
1.0 & \texttt{SXS:BBH:3896} & -0.8 & 0.8 & 2.0 & \texttt{SXS:BBH:0584} & -0.4 & -0.4 & 2.0 & \texttt{RIT:BBH:0767} & 0.95 & 0.95 \\
1.0 & \texttt{SXS:BBH:2423} & 0.85 & 0.85 & 2.0 & \texttt{SXS:BBH:4115} & -0.9 & 0.9 & 2.0 & \texttt{RIT:BBH:0782} & 0.0 & 0.95 \\
1.0 & \texttt{SXS:BBH:0172} & 0.98 & 0.98 & 2.0 & \texttt{SXS:BBH:2107} & -0.87 & 0.85 & 2.0 & \texttt{RIT:BBH:0796} & 0.95 & 0.0 \\
1.0 & \texttt{SXS:BBH:3978} & 0.95 & 0.95 & 2.0 & \texttt{SXS:BBH:0354} & -0.2 & 0.4 & 1.0 & \texttt{GT0420} & 0.2 & 0.2 \\
1.0 & \texttt{SXS:BBH:0329} & -0.8 & -0.8 & 2.0 & \texttt{SXS:BBH:2108} & -0.85 & -0.85 & 1.0 & \texttt{GT0546} & 0.0 & 0.2 \\
1.0 & \texttt{SXS:BBH:3929} & -0.24 & -0.1 & 2.0 & \texttt{SXS:BBH:0334} & -0.8 & -0.8 & 1.0 & \texttt{GT0574} & 0.1 & 0.1 \\
1.0 & \texttt{SXS:BBH:1509} & -0.24 & -0.1 & 2.0 & \texttt{SXS:BBH:0335} & -0.8 & 0.8 & 1.5 & \texttt{GT0479} & 0.2 & 0.2 \\
1.0 & \texttt{SXS:BBH:1506} & 0.46 & -0.32 & 2.0 & \texttt{SXS:BBH:2128} & 0.6 & -0.6 &      &      &      &      \\
1.0 & \texttt{SXS:BBH:3917} & 0.68 & 0.67 & 2.0 & \texttt{SXS:BBH:0574} & 0.4 & 0.4 &      &      &      &      
\end{tabular}
\end{ruledtabular}
\end{table*}

\section{Post-Newtonian expressions for relative mode amplitudes} \label{app:pn_expressions}
The \ac{pn} waveform mode amplitudes relative to the \( (2,2) \) mode up to $\mathcal{O}(v^5)$ are given by \cite{blanchet2024post}, and we rewrite them as $\hat{A}_{\ell m}$ in Eq.~\eqref{eq:pn_amplitudes},
\begin{widetext}
\begin{subequations} \label{eq:pn_amplitudes}
\begin{align}
    \label{eq:nonspinning_pn_amps}
    \hat{A}_{22}^{(\text{PN})} &= 8 \sqrt{\frac{\pi }{5}} \eta  v^2 + \frac{4}{21} \sqrt{\frac{\pi }{5}} \eta  (55 \eta -107) v^4  + \mathcal{O}(v^5)\\
    \hat{A}_{21}^{(\text{PN})}  &=  \frac{\delta  v}{3} +\frac{1}{252} \delta  (163-50 \eta ) v^3-\frac{1}{3} \pi  \delta  v^4 + \mathcal{O}(v^5) \\
    \hat{A}_{33}^{(\text{PN})}  &=  \frac{3}{4} \sqrt{\frac{15}{14}} \delta  v +\frac{1}{56} \sqrt{\frac{15}{14}} \delta  (29 \eta -61) v^3 + \frac{3}{4} \sqrt{\frac{15}{14}} \pi  \delta  v^4 + \mathcal{O}(v^5)\\
    \hat{A}_{32}^{(\text{PN})} &= \frac{1}{3} \sqrt{\frac{5}{7}} (1-3 \eta ) v^2 + \frac{(254-5 \eta  (16 \eta +113)) v^4}{378 \sqrt{35}} + \mathcal{O}(v^5)\\
    \hat{A}_{31}^{(\text{PN})} &= \frac{\delta  v}{12 \sqrt{14}}-\frac{\delta  (83 \eta +5) v^3}{504 \sqrt{14}}-\frac{\pi  \delta  v^4}{12 \sqrt{14}} + \mathcal{O}(v^5)\\
    \hat{A}_{44}^{(\text{PN})} &= -\frac{8}{9} \sqrt{\frac{5}{7}} (3 \eta -1) v^2-\frac{4 (25 \eta  (372 \eta -955)+6568) v^4}{2079 \sqrt{35}} + \mathcal{O}(v^5)\\
    \hat{A}_{43}^{(\text{PN})} &= \frac{9 \delta  (1-2 \eta ) v^3}{4 \sqrt{70}} + \mathcal{O}(v^5)\\
    \hat{A}_{42}^{(\text{PN})} &= \frac{1}{63} \sqrt{5} (1-3 \eta ) v^2 + \frac{\left(7080 \eta ^2+7495 \eta -3292\right) v^4}{29106 \sqrt{5}} + \mathcal{O}(v^5)\\
    \hat{A}_{41}^{(\text{PN})} &= \frac{\delta  (1-2 \eta ) v^3}{84 \sqrt{10}} + \mathcal{O}(v^5)
\end{align}
\end{subequations}
\end{widetext}
where the total mass and luminosity distance have been set to unity to match convention in \ac{nr} simulations.

The spin contributions to the mode amplitudes can be found in Refs.~\cite{henry2022spin, pan2014inspiral, arun2009higher}. We rewrite these relative to the \( (2,2) \) mode up to $\mathcal{O}(v^6)$ in Eq.~\eqref{eq:pn_spin_exp}.
\begin{widetext}
    \begin{subequations}
    \label{eq:pn_spin_exp}
        \begin{align}
            \hat{A}_{22}^{(\text{PN, S})} &= \frac{32}{3} \sqrt{\frac{\pi }{5}} \eta v^5 \left[(\eta -1) \chi _s-\delta  \chi _a\right] \label{eq:22_spin}\\
            \hat{A}_{21}^{(\text{PN, S})} &= -\frac{1}{2}v^2(\chi_a+\delta\chi_s)+\frac{1}{252}v^4\left[(-251+947\eta)\chi_a+251\delta(-1+\eta)\chi_s \right] \notag \\
            & + \frac{1}{6}v^6(\chi_a + \delta\chi_s)\left(3\pi - 4\delta \chi_a + 4(-1+\eta)\chi_s\right) + \mathcal{O}(v^6) \\
            \hat{A}_{32}^{(\text{PN, S})} &= \frac{4}{3}\sqrt{\frac{5}{7}}\left[ \eta  v^3 \chi _s +\frac{1}{84}v^5\left(7 \delta  (31 \eta -4) \chi _a+((81-208 \eta ) \eta -28) \chi _s\right)\right] + \mathcal{O}(v^6) \\
            \hat{A}_{43}^{(\text{PN, S})} &= -\frac{9}{8}\sqrt{\frac{5}{14}}v^4 \eta (\chi_a - \delta \chi_s) + \mathcal{O}(v^6) \\
            \hat{A}_{41}^{(\text{PN, S})} &= - \frac{1}{168}\sqrt{\frac{5}{2}}v^4 \eta (\chi_a - \delta \chi_s) + \mathcal{O}(v^6)
        \end{align}
    \end{subequations}
\end{widetext}

Note that these expressions are given relative to the $(2,2)$ mode, and thus differ from the expressions for individual $(\ell, m)$ mode amplitudes typically reported in the literature.

\section{Aligned-Spin Fit Snapshots} \label{app:alignedspin_snapshots}
\begin{figure*}
    \centering
    \includegraphics[width=0.9\linewidth]{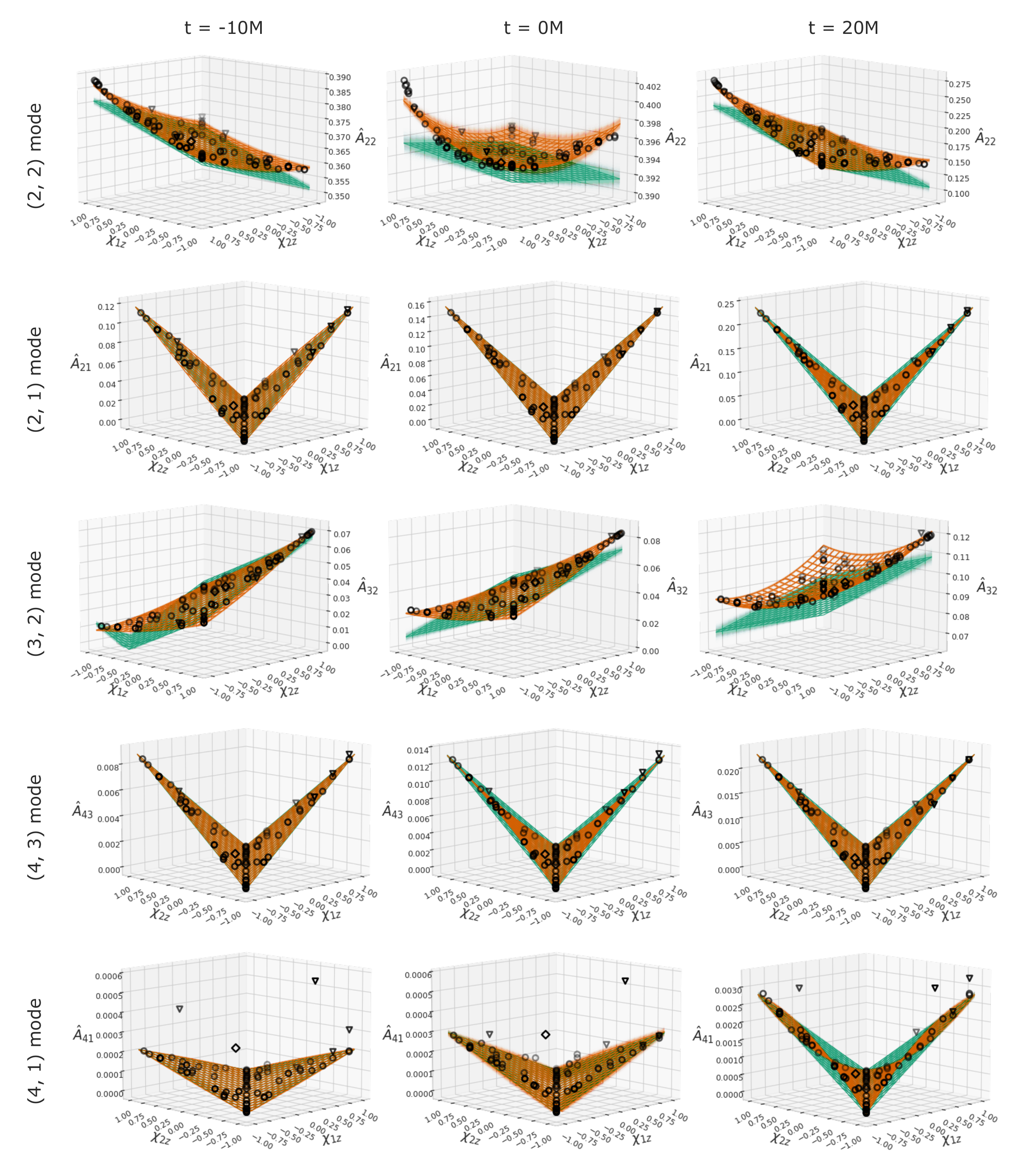}
    \caption{\label{fig:alignedspin_q1.0} Leading-order and quadratic fit snapshots for aligned-spin simulations with $\mathbf{q=1.0}$: Snapshots of the $(\ell, m)$ mode amplitudes, normalized to the $(2,2)$ mode, at selected times $t$ relative to the peak of the $(2,2)$ mode, displayed on a grid of $\chi_{1z}$ and $\chi_{2z}$. Best-fit surfaces are drawn from posteriors based on the leading-order fits (green) defined in Eq.~\eqref{eq:alignedspin_lo_fits} and the quadratic fits (orange) defined in Eq.~\eqref{eq:alignedspin_ho_fits} are shown. Markers denote \ac{nr} data points from the \ac{sxs} (circle), \ac{rit} (triangle), and MAYA (diamond) catalogs. The best-fit surfaces were obtained using data from the \ac{sxs} catalog only. All plots are oriented facing the plane defined by the leading-order dependence on $\chi_s$ and $\chi_a$.}
\end{figure*}

\begin{figure*}
    \centering
    \includegraphics[width=0.9\linewidth]{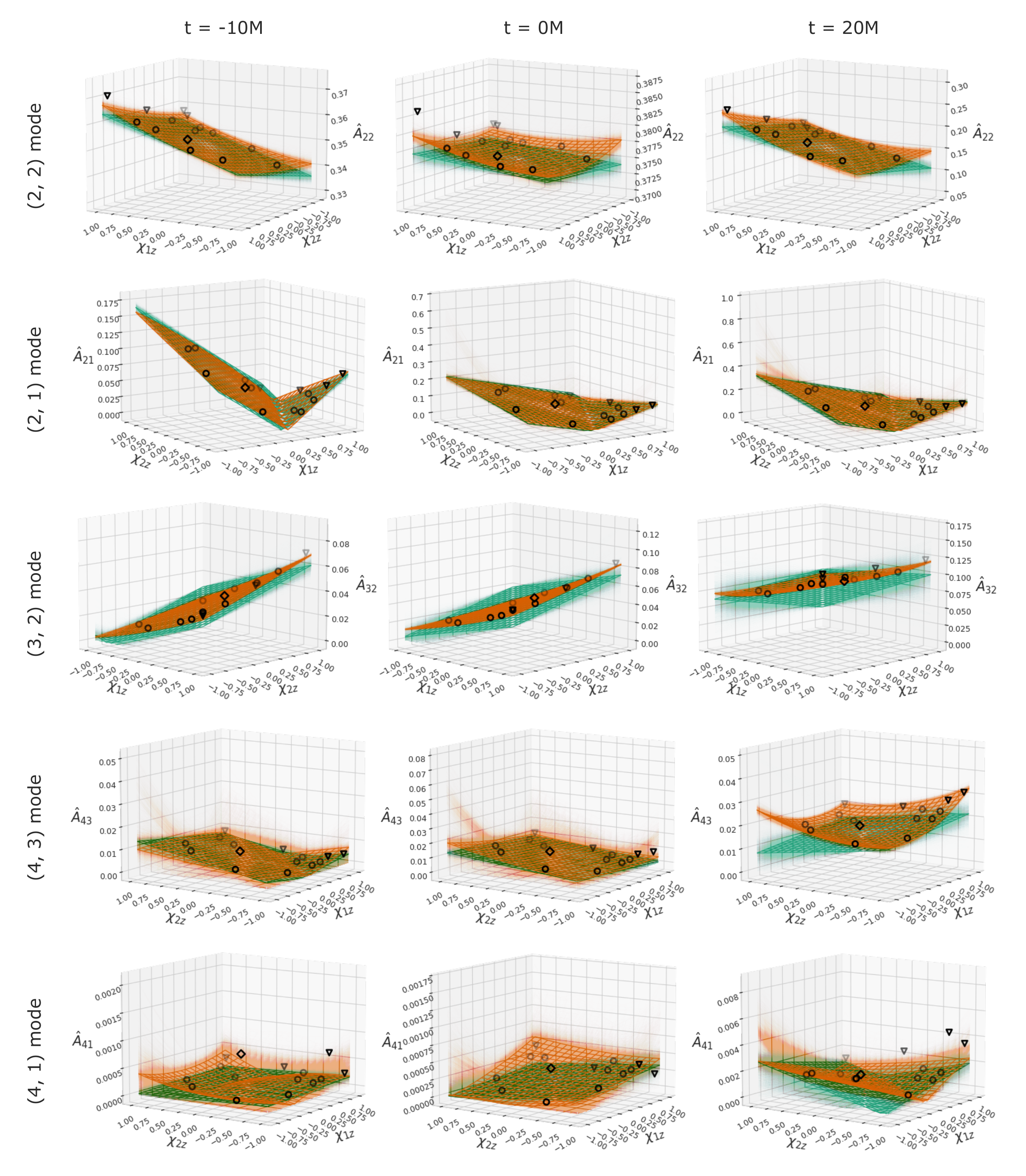}
    \caption{\label{fig:alignedspin_q1.5} Leading-order and quadratic fit snapshots for aligned-spin simulations with $\mathbf{q=1.5}$: Snapshots of the $(\ell, m)$ mode amplitudes, normalized to the $(2,2)$ mode, at selected times $t$ relative to the peak of the $(2,2)$ mode, displayed on a grid of $\chi_{1z}$ and $\chi_{2z}$. Best-fit surfaces are drawn from posteriors based on the leading-order fits (green) defined in Eq.~\eqref{eq:alignedspin_lo_fits} and the quadratic fits (orange) defined in Eq.~\eqref{eq:alignedspin_ho_fits} are shown. The best-fit surfaces were obtained using data from the \ac{sxs} catalog only. Markers denote the \ac{nr} catalog associated with each data point. All plots are oriented facing the plane defined by the leading-order dependence on $\chi_s$ and $\chi_a$.}
\end{figure*}

\begin{figure*}
    \centering
    \includegraphics[width=0.9\linewidth]{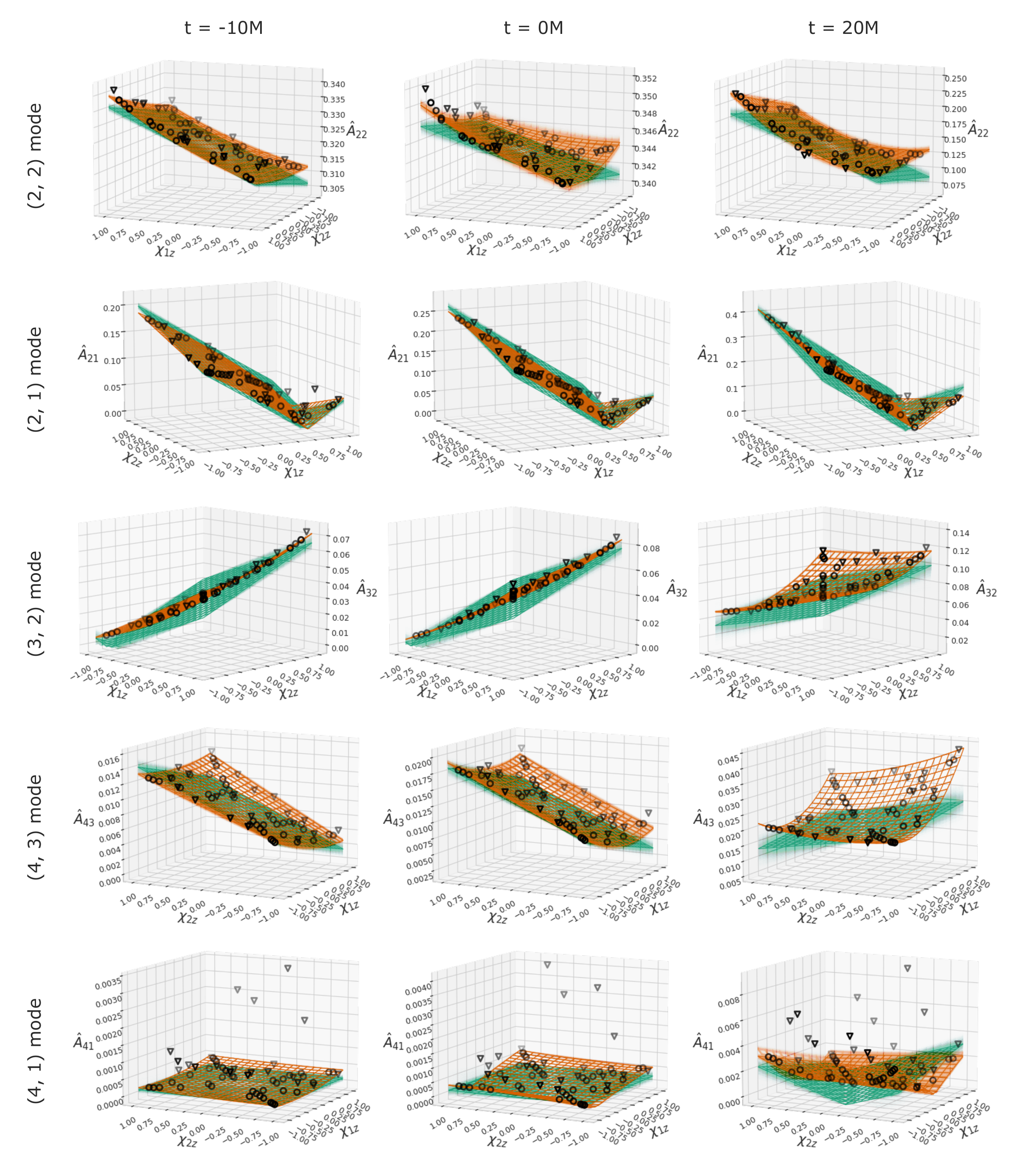}
    \caption{\label{fig:alignedspin_q2.0} Leading-order and quadratic fit snapshots for aligned-spin simulations with $\mathbf{q=2.0}$: Snapshots of the $(\ell, m)$ mode amplitudes, normalized to the $(2,2)$ mode, at selected times $t$ relative to the peak of the $(2,2)$ mode, displayed on a grid of $\chi_{1z}$ and $\chi_{2z}$. Best-fit surfaces are drawn from posteriors based on the leading-order fits (green) defined in Eq.~\eqref{eq:alignedspin_lo_fits} and the quadratic fits (orange) defined in Eq.~\eqref{eq:alignedspin_ho_fits} are shown. The best-fit surfaces were obtained using data from the \ac{sxs} catalog only. Markers denote the \ac{nr} catalog associated with each data point. All plots are oriented facing the plane defined by the leading-order dependence on $\chi_s$ and $\chi_a$.}
\end{figure*}

Figures~\ref{fig:alignedspin_q1.0}--\ref{fig:alignedspin_q2.0} compare the leading-order \ac{pn}-inspired fits and the quadratic fits, omitting the linear fits for simplicity. For these figures, the coefficients are inferred using the Bayesian regression procedure outlined in Appendix~\ref{app:bayesian}, and the fit surfaces are drawn from the resulting posteriors.

\bibliographystyle{apsrev4-2}
\providecommand{\noopsort}[1]{}\providecommand{\singleletter}[1]{#1}%

\end{document}